\documentclass[smallcondensed,natbib]{svjour3}    

\smartqed  

\usepackage{aas_macros}
\usepackage[colorlinks,linkcolor=blue]{hyperref}
\usepackage{graphicx}
\usepackage{amsmath,amssymb}
\usepackage[flushleft]{threeparttable}
\usepackage{array}
\newcolumntype{C}[1]{>{\centering\arraybackslash}p{#1}}
\newcommand{\changed}{}

\begin{document}

\title{Transfer, loss and physical processing of water in hit-and-run collisions of planetary embryos}

\titlerunning{Water transfer and loss in hit-and-run collisions} 

\author{C. Burger         \and
		T. I. Maindl	\and
        C. M. Sch\"afer 
}


\institute{C. Burger \and T. I. Maindl \at 
				Department of Astrophysics, University of Vienna, T\"urkenschanzstra\ss{}e 17, 1180 Vienna \\
              \email{c.burger@univie.ac.at}
              \and C. M. Sch\"afer \at 
              Institut f{\"u}r Astronomie und Astrophysik, Eberhard Karls Universit{\"a}t T{\"u}bingen, Auf der Morgenstelle 10, 72076 T{\"u}bingen
              }

\maketitle

\begin{abstract}
Collisions between large, similar-sized bodies are believed to shape the final characteristics and composition of terrestrial planets. Their inventories of volatiles such as water, are either delivered or at least significantly modified by such events.
Besides the transition from accretion to erosion with increasing impact velocity, similar-sized collisions can also result in hit-and-run outcomes for sufficiently oblique impact angles and large enough projectile-to-target mass ratios.
We study volatile transfer and loss focusing on hit-and-run encounters by means of Smooth Particle Hydrodynamics simulations, including all main parameters: impact velocity, impact angle, mass ratio, and also the total colliding mass.
We find a broad range of overall water losses, up to 75\% in the most energetic hit-and-run events, and confirm the much more severe consequences for the smaller body also for stripping of volatile layers.
Transfer of water between projectile and target inventories is found to be mostly rather inefficient, and final water contents are dominated by pre-collision inventories reduced by impact losses, for similar pre-collision water mass fractions.
Comparison with our numerical results shows that current collision outcome models are not accurate enough to reliably predict these composition changes in hit-and-run events. 
To also account for non-mechanical losses we estimate the amount of collisionally vaporized water over a broad range of masses, and find that these contributions are particularly important in collisions of $\sim$\,Mars-sized bodies, with sufficiently high impact energies, but still relatively low gravity.
Our results clearly indicate that the cumulative effect of several (hit-and-run) collisions can efficiently strip protoplanets of their volatile layers, especially the smaller body, as it might be common e.g.\ for Earth-mass planets in systems with Super-Earths. An accurate model for stripping of volatiles that can be included in future planet formation simulations has to account for the peculiarities of hit-and-run events, and track compositional changes in both large post-collision fragments.

\keywords{hydrodynamics \and methods: numerical \and planets and satellites: formation}

\end{abstract}

\section{Introduction}
Collisions on vastly different size scales are ubiquitous during all stages of planet formation. Varying bulk densities, probably due to different ice mass fractions, of outer solar system satellites and KBOs for example indicate a rich collisional history. Even though there is currently a lot of discussion about whether planetary embryos were formed by planetesimal- followed by oligarchic growth, or rather by rapid accretion of small pebbles \citep{Lambrechts2014_pebble_accretion}, it is well established that the late stages of planet formation -- starting once the gas disk disappears and not enough background material for dynamical friction is left -- are dominated by giant collisions between embryos, with sizes from hundreds to thousands of kilometers. This includes mixing of material that condensed at vastly different orbital radii to some degree \citep{Raymond2014_Terrestrial_formation_at_home_and_abroad}.
Isotopic measurements as well as dynamical arguments point to material from -- or at least isotopically similar to -- the outer asteroid belt as the origin of the bulk of Earth's water. The delivery of volatiles such as water, in our solar system as well as in extrasolar systems, is intimately connected to the long-term evolution of volatiles interacting with the environment in the form of gradual loss of atmospheric constituents \citep[see e.g.][]{Odert2017_Escape_fractionation_of_volatiles} or sublimation of exposed surface ice \citep[e.g.][]{Schorghofer2008_Lifetime_of_ice_on_asteroids}, and to the collisional evolution of volatile-carrying bodies. In this study we deal with the latter.

While impacts of small bodies onto much larger ones are usually mostly accretionary, it is well-known that collisions between objects of comparable size can result in a much more diverse spectrum of outcomes.
\citet{Leinhardt2012_Collisions_between_gravity-dominated_bodies_I} developed a comprehensive analytical model to distinguish different collision outcome regimes. Low-velocity impacts result in accretion of at least some projectile material onto the target, while more energetic collisions often lead to erosion or even disruption. For sufficiently oblique impact angles \citep{Genda2012_Merging_criteria_for_giant_impacts} an additional outcome is possible -- hit-and-run. These grazing encounters are characterized by two large post-collision fragments, originating from the target and the projectile. Therefore hit-and-run is usually defined via an accretion efficiency \citep{Asphaug2009_Growth_evolution_of_asteroids} close to zero, given by $\xi = (M_\mathrm{lf} - M_\mathrm{targ})/M_\mathrm{proj}$, with the mass of the largest post-collision fragment $M_\mathrm{lf}$ \changed{and of the (pre-collision) target and projectile}. This means that the target body neither accretes a lot of projectile material nor is it substantially eroded (cf.~Fig.~\ref{fig:snapshots}).
The hit-and-run regime covers a wider range of impact velocities the more oblique collisions are, but is also a strong function of the projectile-to-target mass ratio, with a higher probability for hit-and-run the more similar-sized impacting bodies are.
Hit-and-run encounters are particularly interesting, but also complex, when it comes to transfer and loss of volatiles. While in the other outcome regimes there is at most one large fragment and usually some amount of orders-of-magnitude less massive debris, changes of the volatile fraction on both large survivors are of interest for hit-and-run, as well as transfer between the initial reservoirs on projectile and target in the course of the impact.
The impact energy is partitioned between the colliding bodies, therefore the smaller one of the colliding pair is more affected, resulting in large-scale deformation and stripping of its outer layers, particularly of volatiles.
While central impacts are dominated by shocks, gravitational stresses become more important in oblique encounters, and the most grazing collisions are almost entirely controlled by tidal forces.
Hit-and-run is also important for shaping the final spin of forming planets \citep{Kokubo2010_Formation_under_realistic_accretion}, and is considered an efficient mantle stripping mechanism to explain for instance Mercury's high bulk density, or the formation of Earth's Moon \citep{Reufer2012_Hit-and-run_Moon_formation}.

Current N-body simulations often include relatively simple models for collisional fragmentation which were developed in recent years \citep{Kokubo2010_Formation_under_realistic_accretion, Genda2012_Merging_criteria_for_giant_impacts, Marcus2010_Icy_Super_Earths_max_water_content, Leinhardt2012_Collisions_between_gravity-dominated_bodies_I, Stewart2012_Collisions_between_gravity-dominated_bodies_II}. These models are applied to compute the basic fragmentation outcome of similar-sized collisions \citep[e.g.][]{Chambers2013_Late-stage_accretion_hit-and-run_fragmentation, Quintana2016_Frequency_giant_impacts_Earth-like_worlds}, and also in studies on compositional evolution and mantle stripping for modeling material fractionation \citep[e.g.][and references therein]{Carter2015_Compositional_evolution_during_accretion}.
\changed{A more direct, but computationally very demanding, approach are simulations that combine the long-term dynamical evolution and the hydrodynamics of individual collisions in hybrid codes \citep{Genda2017_Hybrid_code_Ejection_of_iron-bearing_fragments}.}
\changed{In addition to numerical (SPH) simulations some analytical work on volatile losses by impacts is available, as e.g.\ in \citet{Schlichting2015_Atmospheric_mass_loss_impacts}.}
While most of the mentioned studies preferentially treat refractory elements, a comprehensive model for volatile transfer and loss in similar-sized collisions is still not available. Therefore current work on volatile mixing and water delivery usually assumes perfect conservation of volatile inventories in collisions \citep[e.g.][]{Raymond2004_Dynamical_sims_water_delivery, Izidoro2013_Compound_model_origin_water}, even though such material is especially prone to loss processes, either mechanically, thermally, or in connection with the stellar environment.
The reasons are not only its volatile behavior, but also its preferred location in the outer layers of sufficiently differentiated bodies.
Studies on collisional water loss mostly distinguish merely material that is gravitationally bound to the largest \citep{Canup2006_Water_planet-scale_collisions}, or the most significant \citep{Maindl2014_Fragmentation_of_colliding_planetesimals_with_water, MaindlSchaeferHaghighipourBurger2017_water_transport} post-collision fragments, and usually do neither trace the fate of individual inventories on the two large hit-and-run fragments, nor volatile transfer between projectile and target during the collision.
Often these studies do not consider dependencies on all important parameters, and commonly fix especially the mass ratio of the colliding bodies.
We try to close this gap in this work with a dedicated study on volatile transfer and loss in hit-and-run collisions.
\changed{The obtained results and insights are a necessary basis for the development and inclusion of realistic models for volatile transport in future planet formation simulations.}

We describe the applied numerical methods in Sect.~\ref{sect:methods}, and our set of simulation scenarios in Sect.~\ref{sect:scenarios}.
The results are presented and described in Sect.~\ref{sect:results}, before we discuss them and conclude in Sect.~\ref{sect:discussion_conclusions}.
A description of our relaxation approach, as well as collision outcomes and results for water transfer and loss for all computed scenarios are included in the Appendix.

\section{Methods}
\label{sect:methods}
We performed Smooth Particle Hydrodynamics (SPH) simulations of collisions between similar-sized planetary embryos.
Our SPH hydrocode utilizes GPU hardware to allow for high resolution runs with still reasonable computing times and has been successfully applied to different aspects of collision and impact processes \citep{MaindlDvorakLammer2015_Impact_heating_on_early_Mars, DvorakMaindlBurger2015_Nonlinear_phenomena_complex_systems_article, HaghighipourMaindl2016_Main_belt_comets}.
In the following we briefly summarize only its main features and refer to \citet{Schaefer2016_miluphcuda} for a comprehensive description.
The SPH code includes self-gravity and in addition to hydrodynamic objects it is also capable of modeling full solid-body physics, with a von Mises plasticity model \citep{Benz1994_Impact_sims_with_fracture_I} and a brittle failure/fragmentation model introduced to SPH by \citet{Benz1995_Sims_brittle_solids_SPH}.
This model is based on a Weibull distribution of flaws, with parameters from \citet{Nakamura2007_Weibull_parameters_basalt} for basalt and from \citet{Lange1984_Weibull_parameters_water_ice} for water ice.
\changed{The von Mises yield criterion does not consider the pressure-dependence of shear strength, and is therefore not the ideal choice for geologic materials, where shear strength is in general a complex function of pressure, and to a lesser degree also of temperature, strain, and even strain-rate. These issues are discussed further in Sect.~\ref{sect:material_strength}.}

\changed{To overcome the problems associated with different particle masses of the standard SPH method we apply the modified SPH approach by \citet{Ott2003_modified_SPH_for_large_density_differences} in all simulations.}

The thermodynamical behavior is modeled by means of the non-linear \citet{Tillotson1962_Metallic_EOS} equation of state (eos), applicable over a wide range of physical conditions \citep[see also][]{Melosh1989_Impact_cratering}.
The Tillotson eos has two distinct analytical forms, covering different regions of density $\varrho$ and (specific) internal energy $e$. For compressed regions ($\varrho \geq \varrho_0$) and cold expanded states ($\varrho < \varrho_0$; $e < e_\mathrm{iv}$ -- the energy of incipient vaporization) it reads
\begin{equation}
p\,(\varrho,e) = \left[ a + \frac{b}{1+e/(e_0 \eta^2)} \right]\varrho e + A\mu + B\mu^2\ ,
\label{eq:till_1}
\end{equation}
with $\eta = \varrho / \varrho_0$ and $\mu = \eta-1$.
Expanded and vaporized states ($\varrho < \varrho_0$; $e \geq e_\mathrm{cv}$ -- the energy of complete vaporization) are described by
\begin{equation}
p\,(\varrho,e) = a\varrho e + \left[ \frac{b\varrho e}{1+e/(e_0 \eta^2)} \right. + \left. A\mu \exp \left\{-\beta \left(\frac{\varrho_0}{\varrho}-1\right)\right\} \right] \times \exp \left\{-\alpha\left(\frac{\varrho_0}{\varrho}-1\right)^2\right\}\ .
\label{eq:till_2}
\end{equation}
In the partial vaporization regime ($e_\mathrm{iv} < e < e_\mathrm{cv}$) a weighted average of (\ref{eq:till_1}) and (\ref{eq:till_2}) is used to interpolate.
\changed{For cold expanded states ($e<e_\mathrm{cv}$ and $\varrho<\varrho_0$) a low-density pressure cutoff is applied by setting $p=0$ for $\varrho/\varrho_0<0.9$ to avoid unphysical tension in states where the material rather forms droplets or fractures instead of remaining a continuum.}
The required material parameters for $\varrho_0$, $e_0$, $e_\mathrm{iv}$, $e_\mathrm{cv}$, $A$, $B$, $a$, $b$, $\alpha$, $\beta$ for basalt and water ice are from \citet{Benz1999_Catastrophic_disruptions_revisited}, and those for iron from \citet{Melosh1989_Impact_cratering}.
To estimate the amount of vaporized material after a collision, we define vaporization as $e \geq e_\mathrm{cv}$ and $\varrho < \varrho_0$, i.e.\ falling into region (\ref{eq:till_2}) of the Tillotson eos. For water these values are $\varrho_0 = 917\,\mathrm{kg}/\mathrm{m}^3$ and $e_\mathrm{cv} = 3.04$\,MJ/kg.
\changed{However, the water vapor fractions derived by this method should be considered rather qualitative estimates, while for a robust quantitative assessment other, more complex eos like ANEOS \citep[see e.g.][]{Melosh2007_A_hydrocode_EOS_for_SiO2} are required. While ANEOS would provide a thermodynamically consistent treatment of phase changes, it comes with other drawbacks, like a requirement for higher resolution \citep{Reinhardt2017_Numerical_aspects_of_giant_impact}.}

The simulated bodies were initialized with relaxed internal structures following self-consistent, semi-analytically computed hydrostatic profiles, with internal energy values from adiabatic compression. 
\changed{The details of this relaxation approach are summarized in Appendix A.}
In the initial configuration the colliding bodies are placed apart at a distance of several times the sum of their radii to allow for build-up of tidal effects, and are sent on analytical two-body orbits that lead to the desired collision parameters.
All simulations were computed until a final state, characterized by a large distance of the major fragments (typically dozens of times the sum of projectile and target radius). The identification of the final fragments is then done as a post-processing step.
First, clumps of spatially connected SPH particles are identified by a friends-of-friends algorithm. Then we iteratively search for gravitationally bound aggregates of such clumps, where we first compute a seed, consisting of the most massive clump and all other clumps that are mutually gravitationally bound to it. This aggregate of clumps (its barycenter) serves as the starting point of an iterative procedure, where we go through the list of clumps, check whether they are gravitationally bound to it, and add/remove them if necessary from the aggregate of clumps. Once this procedure converges we identify the result as the largest fragment, and repeat it once again to identify the second largest fragment as well, by starting with the largest still unbound clump (clumps that are already included in the largest fragment are always left there).
\changed{Convergence towards a self-consistent final state is usually achieved after few iterations. A variety of similar methods is commonly applied to identify the largest collisional fragments, which either include a preceding friends-of-friends step \citep[e.g.][]{Genda2012_Merging_criteria_for_giant_impacts,Benz1999_Catastrophic_disruptions_revisited}, or start identifying gravitationally bound mass directly on the individual particle level \citep[as e.g.\ in][]{Movshovitz2016_Impact_disruption_new_simulation_data_and_scaling,Asphaug2010_Similar_sized_collisions_diversity_of_planets,Marcus2009_Collisional_stripping_of_Super-Earths}, or apply both approaches \citep[e.g.][]{Genda2015_Resolution_dependence_of_collisions}.}
In a hit-and-run collision the two largest fragments usually originate from the target and the projectile, respectively.

\section{Scenarios}
\label{sect:scenarios}
	\begin{figure}
	\centering
	\includegraphics[width=0.45\hsize]{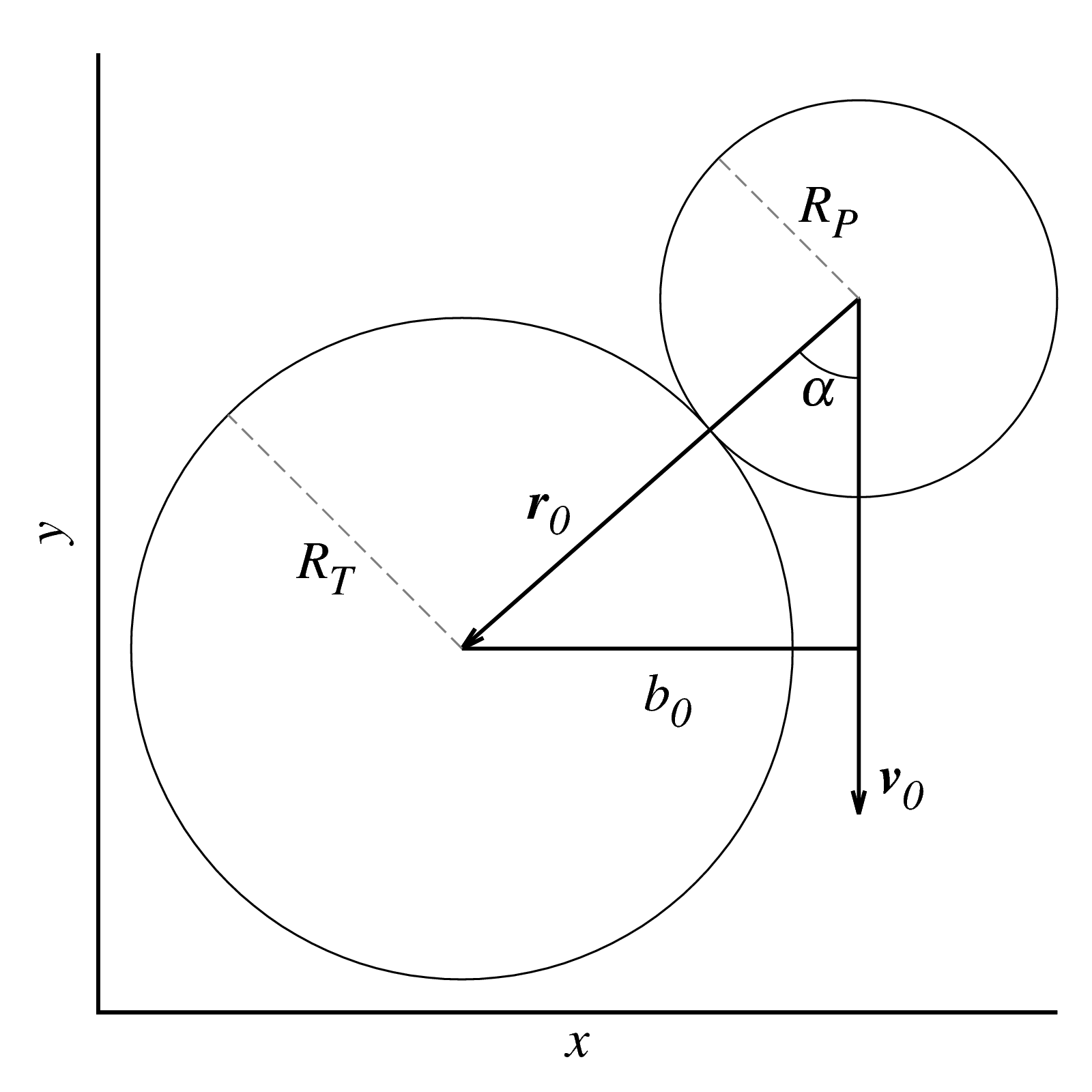}
	\includegraphics[width=0.45\hsize]{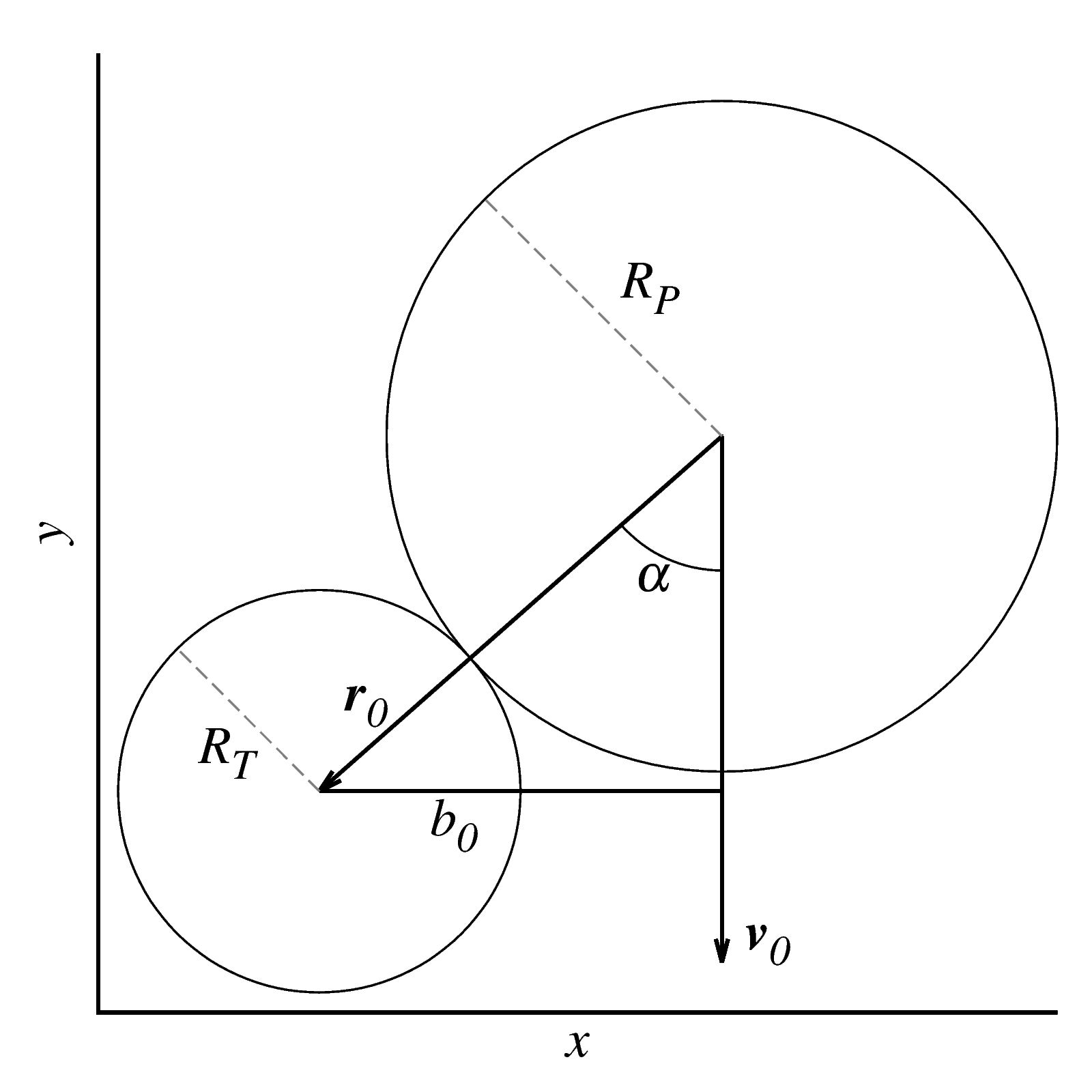}
	\caption{Collision geometry in a frame centered on the target, with the conventional definition that the target is the larger body (left) and also from the perspective of the smaller one of the colliding pair (right). The impact angle $\alpha$ spans values up to $90^\circ$ and is $0^\circ$ for a head-on collision. The impact velocity vector is labelled $\mathbf{v}_0$, where the actual impact velocity is $v_0 = |\mathbf{v}_0|$.}
	\label{fig:collision_geometry}
	\end{figure}
Our scenarios comprise collisions between differentiated, embryo-sized bodies. Most of them have a 3-layered structure, consisting of an iron core (25\% by mass), a silicate basalt mantle (65\%), and a water (ice) shell (10\%).
Including volatile inventories in both colliding bodies allows us to track transfer from the projectile to the target and vice versa, as well as losses from both bodies, from the results of a single simulation run, by tracing the respective SPH particles (target vs.\ projectile water). Therefore this eliminates the necessity to run an otherwise similar collision twice, once with a water-covered target hit by a dry projectile and once the other way around (see~Sect.~\ref{sect:water_amount_distribution}).
Even with the composition fixed the parameter space is large, where the impact velocity in units of the mutual escape velocity\footnote{The two-body escape velocity is given by $v_\mathrm{esc} = \sqrt{2\,G\,M/r}$ with the gravitational constant $G$, the combined mass $M = m_1+m_2$ and the bodies' separation $r$.} $v/v_\mathrm{esc}$, the impact angle $\alpha$ and the bodies' mass ratio $\gamma = M_\mathrm{proj}/M_\mathrm{targ}$ are the most important ones.
Figure~\ref{fig:collision_geometry} illustrates the definition of the impact velocity and angle, both at the moment of first contact, assuming spherical objects.
While the total colliding mass $M_\mathrm{tot}$ is not of crucial importance to some aspects of giant collisions \citep{Asphaug2010_Similar_sized_collisions_diversity_of_planets, Genda2012_Merging_criteria_for_giant_impacts}, meaning that outcomes are scale invariant, recent results have shown that this holds only limited for transfer and loss of volatiles \citep{Burger2017_hydrodynamic_scaling}.

The range of parameters we chose for the simulation runs is focused on the region in parameter space occupied by hit-and-run collisions, and is also typical for an active planet formation environment.
We have not covered our whole parameter space uniformly, but started with its central region and performed exploratory simulations in various directions we deemed promising for gaining further insights (see~Fig.~\ref{fig:fancy_summary_m1e23}). This was mainly done to reduce computational costs and the effort for post-processing in parameter regions where no new results were expected.
The impact velocity varies between 1.5 and 5.5\,$\times\,v_\mathrm{esc}$. The impact angles include some head-on scenarios for comparison, but are otherwise rather grazing as is typical for hit-and-run events, and range from $30^\circ$ to $90^\circ$. Mass ratios vary from 1:2 up to 1:50 according to the traditional definition that the target is always the larger body, but from the perspective of a single body which is hit by an arbitrarily large projectile the range of mass ratios encompasses values between 1:50 and 50:1 (cf.~Fig.~\ref{fig:one_body_perspective}).
The total mass is $10^{23}$\,kg ($\sim$\,Moon-mass) in most scenarios, representing typical planetary embryos, but was also varied between $10^{22}$ ($\sim$\,1/10 Moon-mass) and $10^{25}$\,kg ($\sim$\,Earth-mass), to study the dependence on $M_\mathrm{tot}$ and thus impact energy, especially in the context of water vapor production.
Most simulations were computed with 100k SPH particles, which was confirmed to be sufficient for our purposes by similar results of selected scenarios, which were run again \changed{with several higher resolutions up to 2.25\,million particles} (Sect.~\ref{sect:resolution}).
In the majority of scenarios the objects are purely hydrodynamical, which is common in this mass-range and justified by the dominance of gravitational stresses over material strength.
However, in order to clarify the possible influence of material strength we compare them also to results obtained with our solid-body material model (Sect.~\ref{sect:material_strength}).

\section{Results}
\label{sect:results}
\begin{figure}
	\centering
	\includegraphics[width=\hsize]{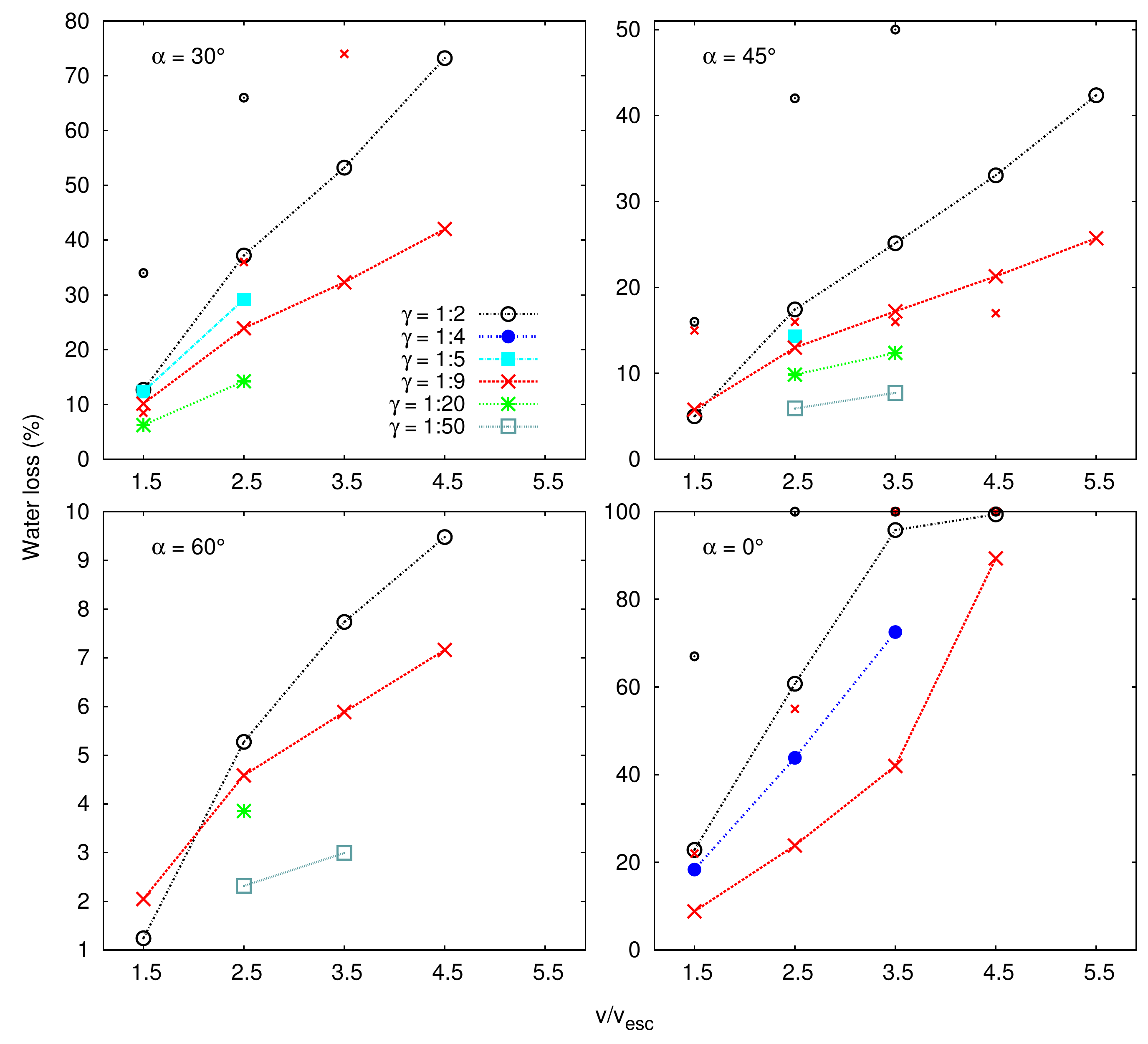}
	\caption{Water loss, defined as escaping the two largest fragments, as a function of impact velocity, for different impact angles $\alpha$ and mass-ratios $\gamma$. Connected points \changed{and single large symbols} are our numerical results, and not connected \changed{small} ones are predictions of the model from \citet[][see~Sect.~\ref{sect:outcome_model}]{Leinhardt2012_Collisions_between_gravity-dominated_bodies_I}. Note the differences in the y-axis range.}
	\label{fig:water_loss_over_vesc}
	\end{figure}

\begin{figure}
	\centering
	\includegraphics[width=\hsize]{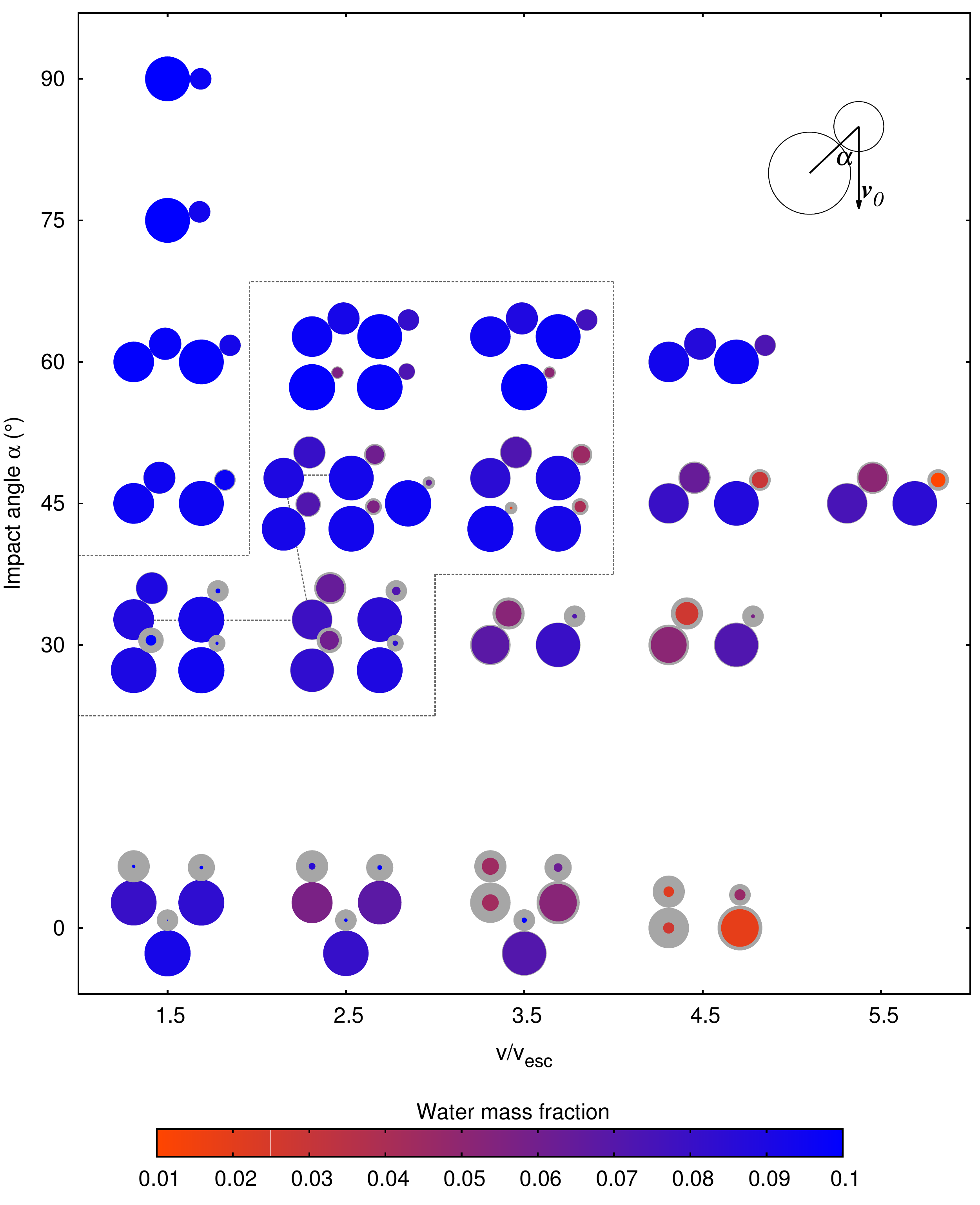}
	\caption{Summary of the $M_\mathrm{tot} = 10^{23}$\,kg scenarios. The pictograms illustrate the collision geometry, the sizes of projectile and target before the collision (light gray) and afterwards (color), as well as the remaining post collision wmf, which are initially $0.1$ (see the online version for color). Circle sizes are proportional to $\mathrm{mass}^{1/3}$, assuming a constant bulk density. The enclosed rectangular region is further examined from a one-body perspective in Fig.~\ref{fig:one_body_perspective}, and the connecting line refers to the scenarios shown in Fig.~\ref{fig:snapshots}.}
	\label{fig:fancy_summary_m1e23}
	\end{figure}

The results of all simulated scenarios are summarized in Tab.~\ref{tab:hit-and-run_results} and Tab.~\ref{tab:head-on} in Appendix B.
In recent studies on collisional water loss results are often presented as the escaping fraction of the colliding bodies initial water inventory, either w.r.t.\ the two largest fragments in case of a hit-and-run \citep[e.g.][]{MaindlSchaeferHaghighipourBurger2017_water_transport}, all significant\footnote{Determined by some mass/SPH-particle-number threshold.} fragments \citep{Maindl2014_Fragmentation_of_colliding_planetesimals_with_water}, or considering only the largest one at all, regardless of the actual collision outcome \citep[e.g.][]{Canup2006_Water_planet-scale_collisions}.
Especially the latter can result in misleading conclusions, since the smaller body in a hit-and-run encounter can carry large amounts of volatiles with it, which are then counted as \emph{lost}, even though they are still part of a large body, possibly similar in size to the largest (Fig.~\ref{fig:snapshots}).
In order to connect to these studies and summarize results we provide a similar plot in Fig.~\ref{fig:water_loss_over_vesc}, where \emph{water loss} refers to the fraction of water bound to neither of the two largest fragments after the collision\footnote{This means that the fraction of lost water is defined as $f_\mathrm{w,lost} = 1-m_\mathrm{w,bound}/m_\mathrm{w,tot}$, where $m_\mathrm{w,tot}$ is the system's total water inventory.}.
However, we do not think that this is a good representation for studying transfer and loss of volatiles in hit-and-run collisions for two reasons, (1) because it does not contain any individual information on the two large post-collision fragments, and (2) because the amount of lost volatiles relative to the pre-collision inventory gives only an approximate figure of the fragments' actual post-collision water mass fractions\footnote{Defined simply as $m_\mathrm{w,frag}/m_\mathrm{frag}$.} (wmf), since they can loose (or gain) other material as well in the process.
To avoid these issues we will discriminate the fate of the two largest hit-and-run fragments in the rest of this study, and furthermore state results as changes in wmf -- relative to overall fragment masses.
We believe this measure is more instructive for tracking the compositional evolution of growing planets, than the change relative to the initial (pre-collision) water inventory.
Figure~\ref{fig:fancy_summary_m1e23} shows an illustration of the bulk of our results following these conventions.

While vaporization (mainly due to shocks) of significant amounts of refractory material happens only in the most energetic events, the situation is different for volatiles, owing to their much lower vaporization energy. Once vaporized, and perhaps heated to high temperatures, this material can escape either thermally, or due to interaction with the (stellar) environment, driven mainly by extreme ultraviolet (EUV) radiation of the young star.
\citet{Odert2017_Escape_fractionation_of_volatiles} studied the loss of steam atmospheres, which is a complicated function of many factors, like the masses of the body and the atmosphere, its thermal state, the distance to the star and its activity, and the frequency of impacts \changed{\citep[see also][]{Schlichting2015_Atmospheric_mass_loss_impacts}}.
Due to the high uncertainties associated with this large number of influences \changed{(and with the determination of the vapor fraction, cf.~Sect.~\ref{sect:methods})} we do not directly include loss of vaporized material in the presented post-collision water contents, but rather estimate the amount of vaporized water for some selected scenarios in a separate Section (\ref{sect:water_vapor}) and discuss the implications thereof. Thus the resulting water contents presented here include all material that is gravitationally bound to the respective fragments, independent of its actual physical state.

\subsection{Water transfer and loss}
\label{sect:water_transfer_and_loss}
	\begin{figure}
	\centering
	\includegraphics[width=0.495\hsize]{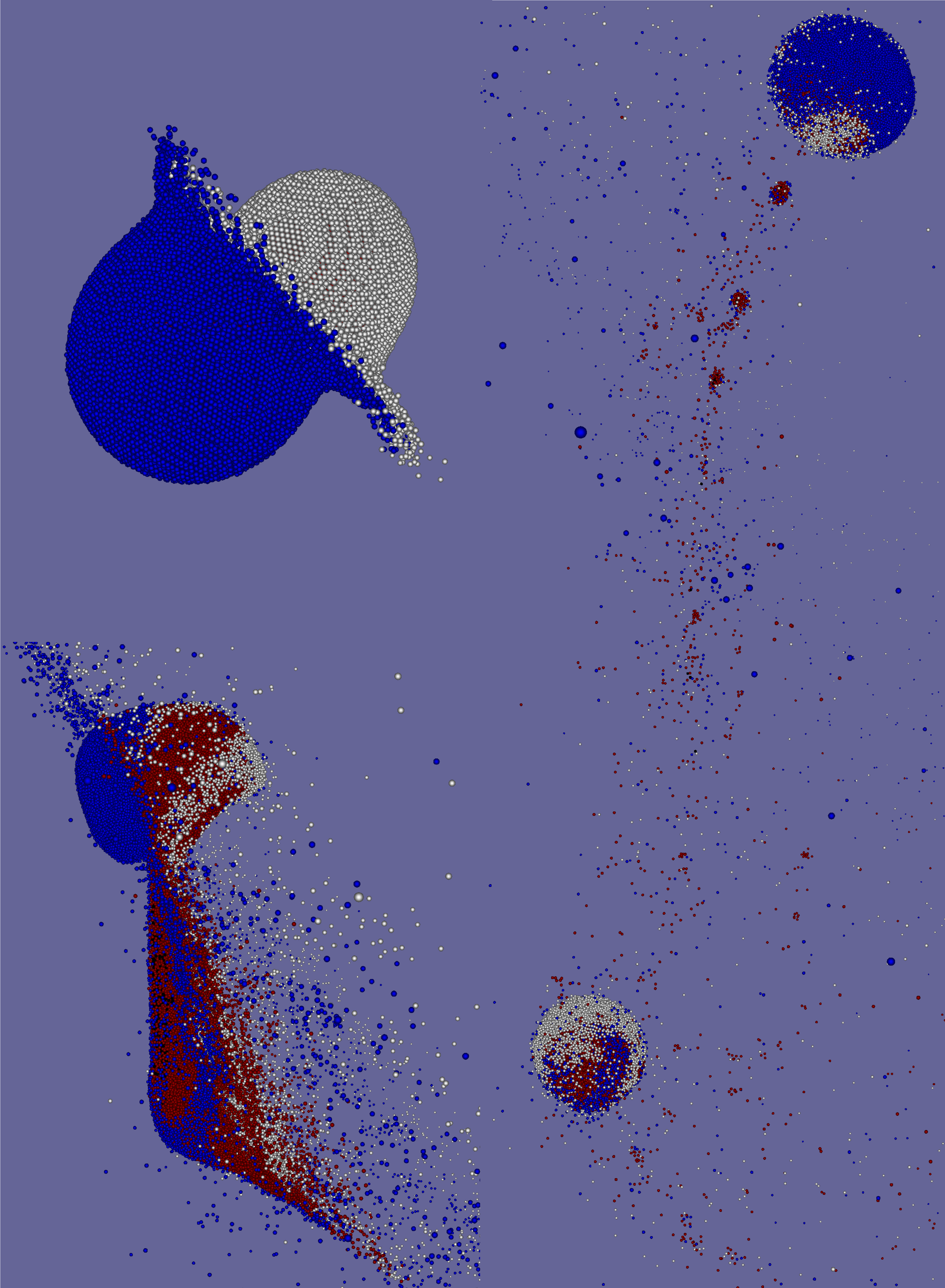}
	\includegraphics[width=0.495\hsize]{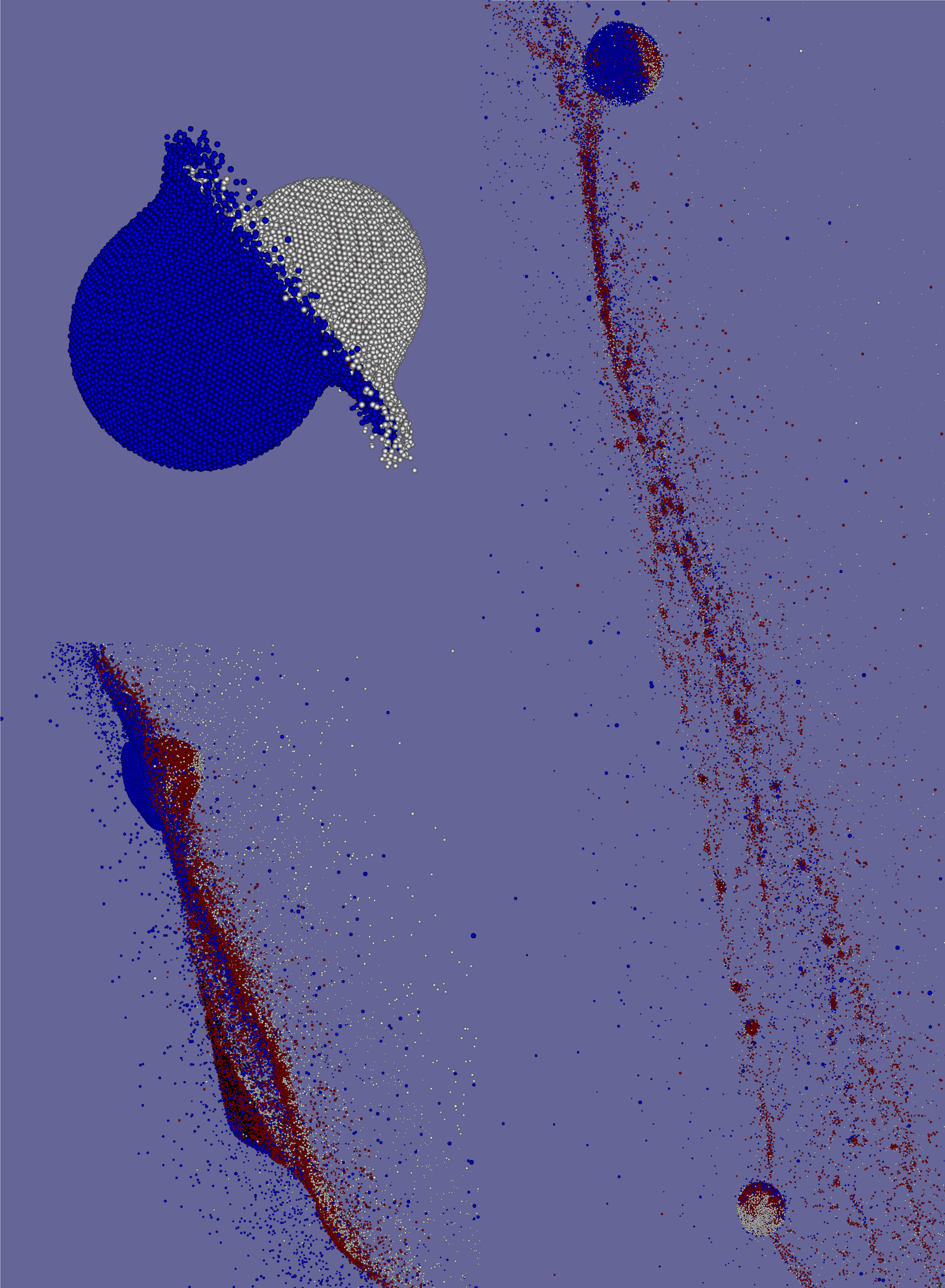} \\
	\vspace{0.5mm}
	\includegraphics[width=0.495\hsize]{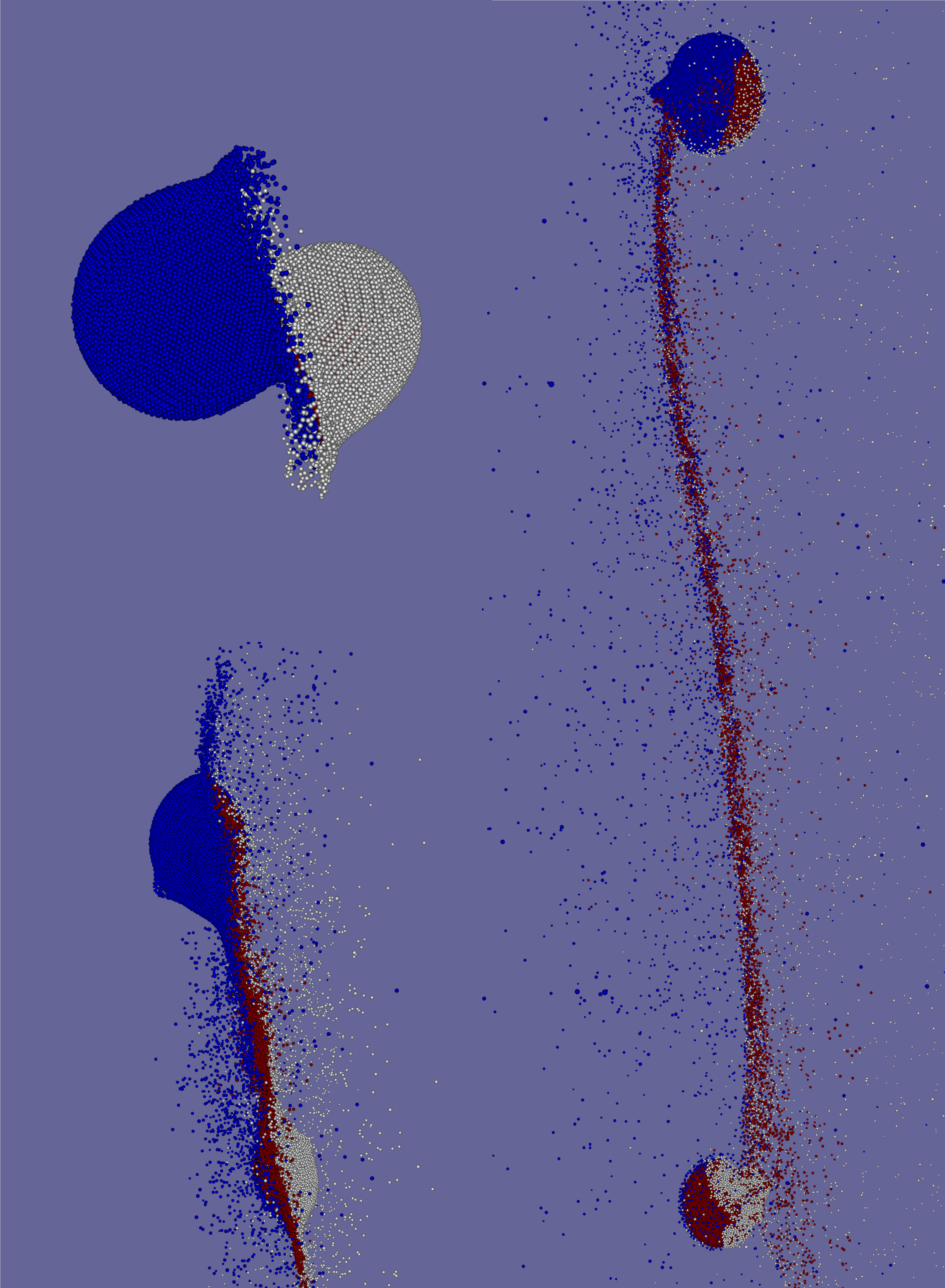}
	\includegraphics[width=0.495\hsize]{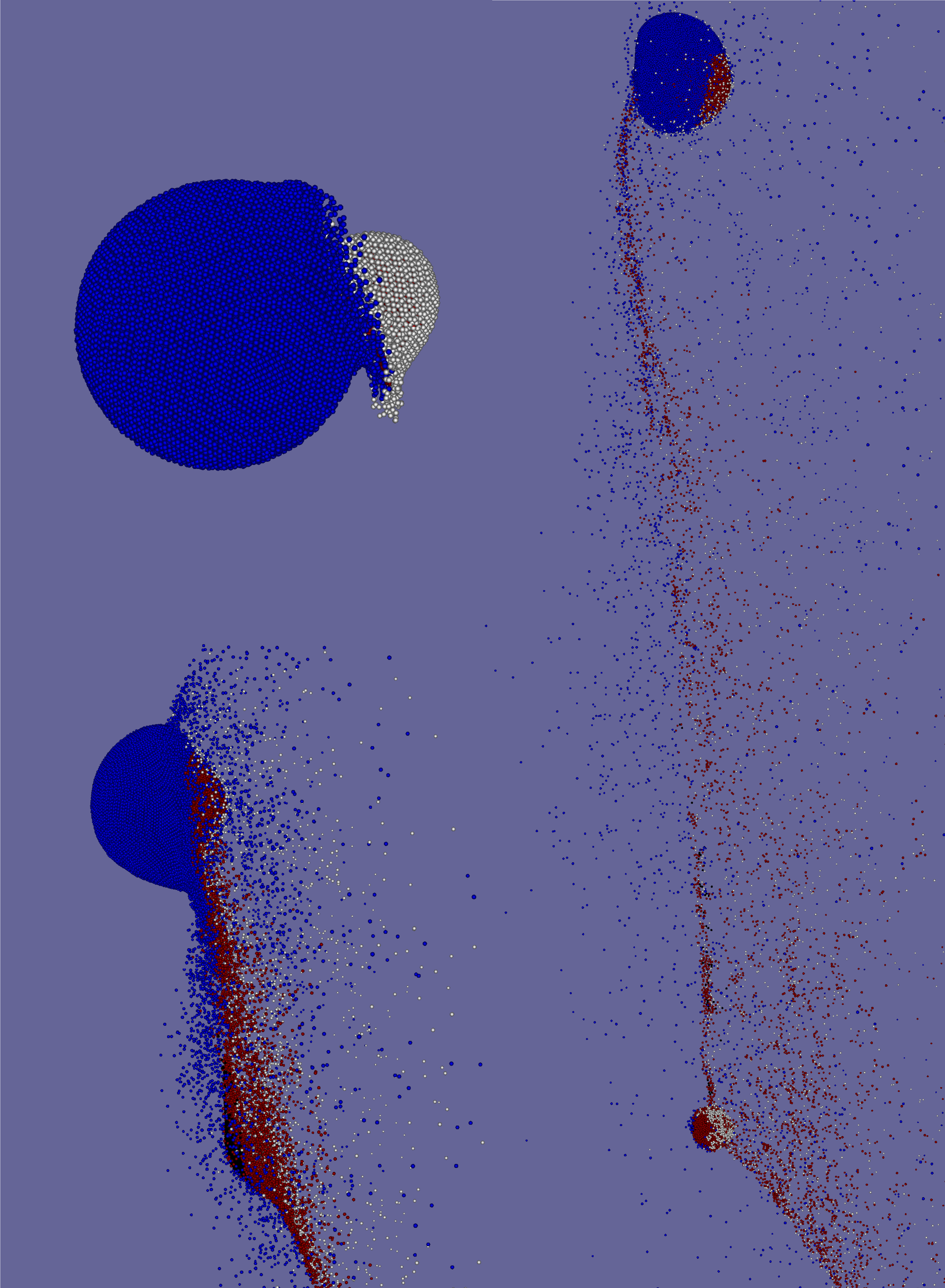}
	\caption{Simulation snapshots illustrating water transfer and loss. Water originally on the target is blue and those initially on the projectile is white. Red and black particles are basalt and iron, respectively (see the online version for color). The chosen scenarios illustrate the impact of collision parameter changes, starting from $v/v_\mathrm{esc} = 1.5$, $\alpha = 30^\circ$, $\gamma$ = 1:2 (upper left), towards a higher velocity of $v/v_\mathrm{esc} = 2.5$ (upper right), towards (additionally) a larger impact angle of $45^\circ$ (lower left), to (additionally) a smaller mass ratio of 1:9 (lower right). The connecting line in Fig.~\ref{fig:fancy_summary_m1e23} illustrates this path in parameter space.}
	\label{fig:snapshots}
	\end{figure}
Due to the high dimensionality of the parameter space most studies on volatile losses in similar-sized collisions hold one or more parameters constant or almost constant, where especially the mass ratio is often fixed to a single value. While we focus on hit-and-run collisions in this study, our scenarios cover a significant range of all basic parameters, thus we can comprehensively investigate their influence on post-collision volatile inventories.
The general increase in volatile losses with impact velocity (Fig.~\ref{fig:water_loss_over_vesc}) is not surprising, but the dependencies on impact angle and mass ratio are more interesting.
There is a strong trend towards losses decreasing with increasing impact angle. While for head-on impacts water losses are considerable even for low $v/v_\mathrm{esc}$, and quickly rise above 50\% for higher velocities, this figure changes drastically for more oblique collisions and for $\alpha \gtrsim 60^\circ$ water losses are small ($<$10\%) even for high-velocity collisions.
For $\alpha = 45^\circ$ they never rise above 50\% for velocities up to $5\times v_\mathrm{esc}$.
These results for overall water loss are in good agreement with previous ones by \citet{Maindl2014_Fragmentation_of_colliding_planetesimals_with_water, MaindlSchaeferHaghighipourBurger2017_water_transport}
who found also a significant decrease of water losses towards oblique impact angles, but studied only fixed mass ratios.
There is also broad agreement to earlier resuts from \citet{Canup2006_Water_planet-scale_collisions}, when their definition of water loss as everything that is not bound to the largest fragment is taken into account. However, their treatment ignores that the impactor in a hit-and-run collision also escapes the target's gravity more or less intact, and can still contain a large volatile reservoir which could be delivered to the same or nearby objects later. 
\citet{Reufer2013_Stripping_in_hit-and-run} focus on mantle stripping of the second largest fragment and consider variations in the four probably most important parameters ($v/v_\mathrm{esc}$, $\alpha$, $\gamma$, $M_\mathrm{tot}$) to some degree.
Our results for mass ratios between 1:50 and 1:2 show a clear increase in combined water loss for more equally-sized bodies, which seems to be more pronounced for lower impact angles (Fig.~\ref{fig:water_loss_over_vesc}).
These large differences can be explained by the fact that the specific energy of a collision $Q_R$ (see Sect.~\ref{sect:water_vapor}) is also a function of $\gamma$\footnote{For the simplifying assumption of equal bulk density $\varrho$ the specific collision energy (Eq.~\ref{eq:Q_R}) can be expressed as \begin{equation}
Q_R = G \left(\frac{4\pi\varrho}{3}\right)^{1/3} \left(\frac{v}{v_\mathrm{esc}}\right)^2 M_\mathrm{tot}^{2/3} \frac{\gamma}{(\gamma+1)^{5/3} (\gamma^{1/3}+1)}\ ,
\end{equation} which increases with $\gamma$ (for all other parameters equal).\label{footnote:Q_R_of_gamma}}, and is roughly doubled when going from $\gamma$ = 1:9 to 1:2 (cf.~Fig.~\ref{fig:vaporization}).
In addition geometry comes into play, because the smaller the projectile the more it is affected relative to the target, but small projectiles have only small water inventories to loose (for an initially equal wmf), while the target retains much of its volatile inventory.

For collisions in the hit-and-run regime however, not only the combined volatile losses, but rather individual ones, as well as transfer between the colliding bodies are of interest.
Figure~\ref{fig:fancy_summary_m1e23} illustrates outcomes for the $M_\mathrm{tot} = 10^{23}$\,kg scenarios and shows both bodies before (light gray circles and impact geometry) and after the collision (colored circles, see online version).
Most interesting in this plot are the big differences between the larger and the smaller one of the colliding pair.
The target barely looses neither mass nor large amounts of water, except for low impact angles and high velocities, and even then only if the projectile has almost the same size.
Disruption of a larger object by the impact of a smaller one was found to be very difficult once bodies grow large enough \citep[e.g.][]{Asphaug2010_Similar_sized_collisions_diversity_of_planets}, and indeed in our scenario set target disruption happens only in high-velocity head-on collisions with large impactors.
The projectile on the other hand is typically much more affected. Except for slow and oblique scenarios it looses considerable amounts of mass, and especially of water, often above 50\% and in some rather high-velocity impacts up to 90\%.
In addition we also did some exploratory simulations with $\alpha = 75^\circ$ and $90^\circ$, which can already be considered a tidal collision, but water losses in these cases are negligible.
Our results also confirm those of earlier studies \citep[e.g.][]{Marcus2010_Icy_Super_Earths_max_water_content}, that the volatile fraction practically never increases in giant collisions between similarly composed objects, but always decreases for both bodies.

	\begin{figure}
	\centering
	\includegraphics[width=\hsize]{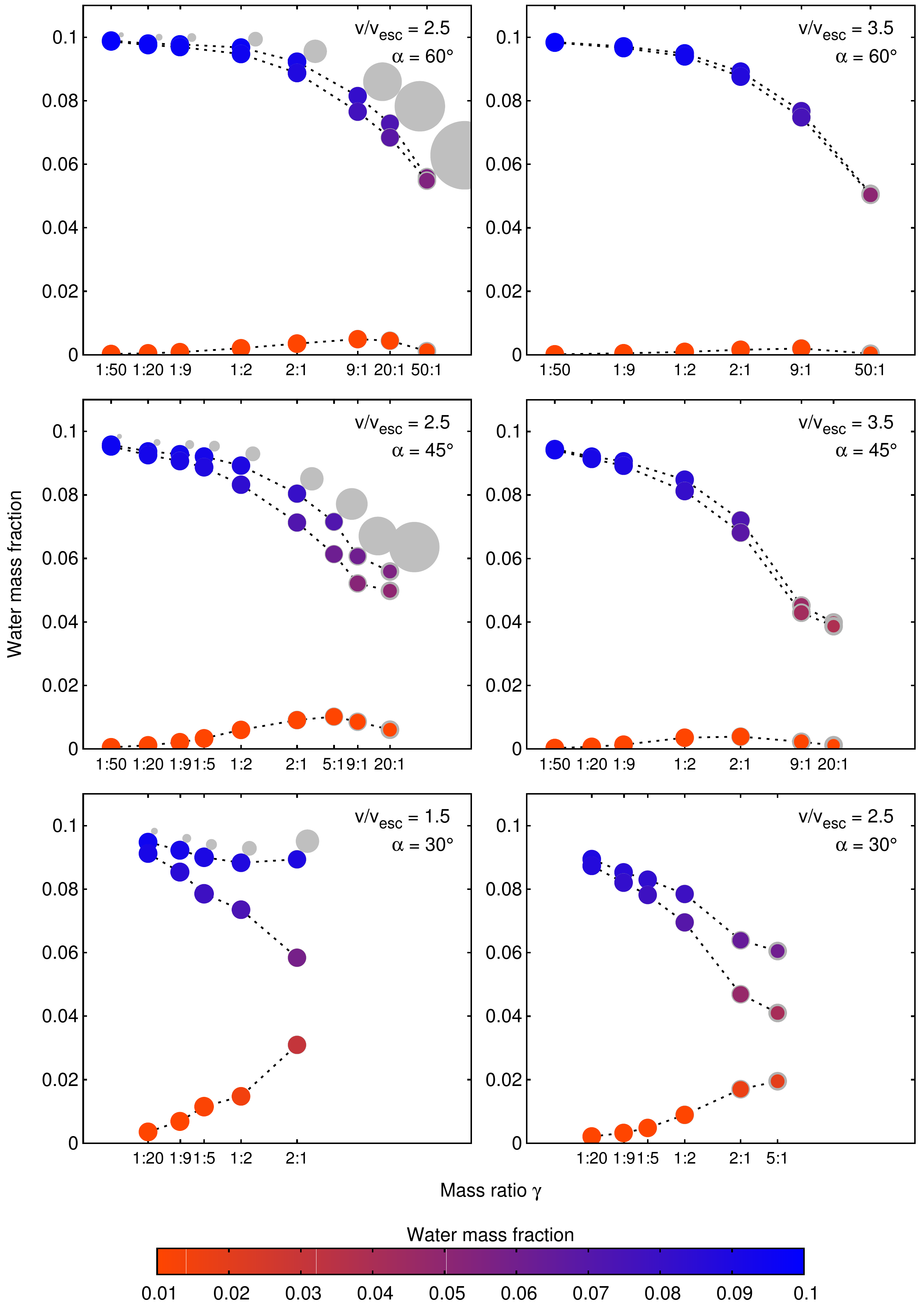}
	\caption{One-body perspective as a function of mass ratio for six different parameter pairs of $v/v_{esc}$ and impact angle $\alpha$ (cf.~Fig.~\ref{fig:fancy_summary_m1e23}). The sizes (grey before and color after the collision, $\propto \mathrm{mass}^{1/3}$) and color-coding indicate what happens to an individual body when hit by a projectile depicted by the light gray circles (see the online version for color). The uppermost graph in each panel represents a collision of equally-composed bodies (wmf = 0.1), the middle graph shows the outcome if only the target body contained water initially, and the lower one depicts the outcome for a dry target.}
	\label{fig:one_body_perspective}
	\end{figure}
\begin{figure}
	\centering
	\includegraphics[width=0.65\hsize]{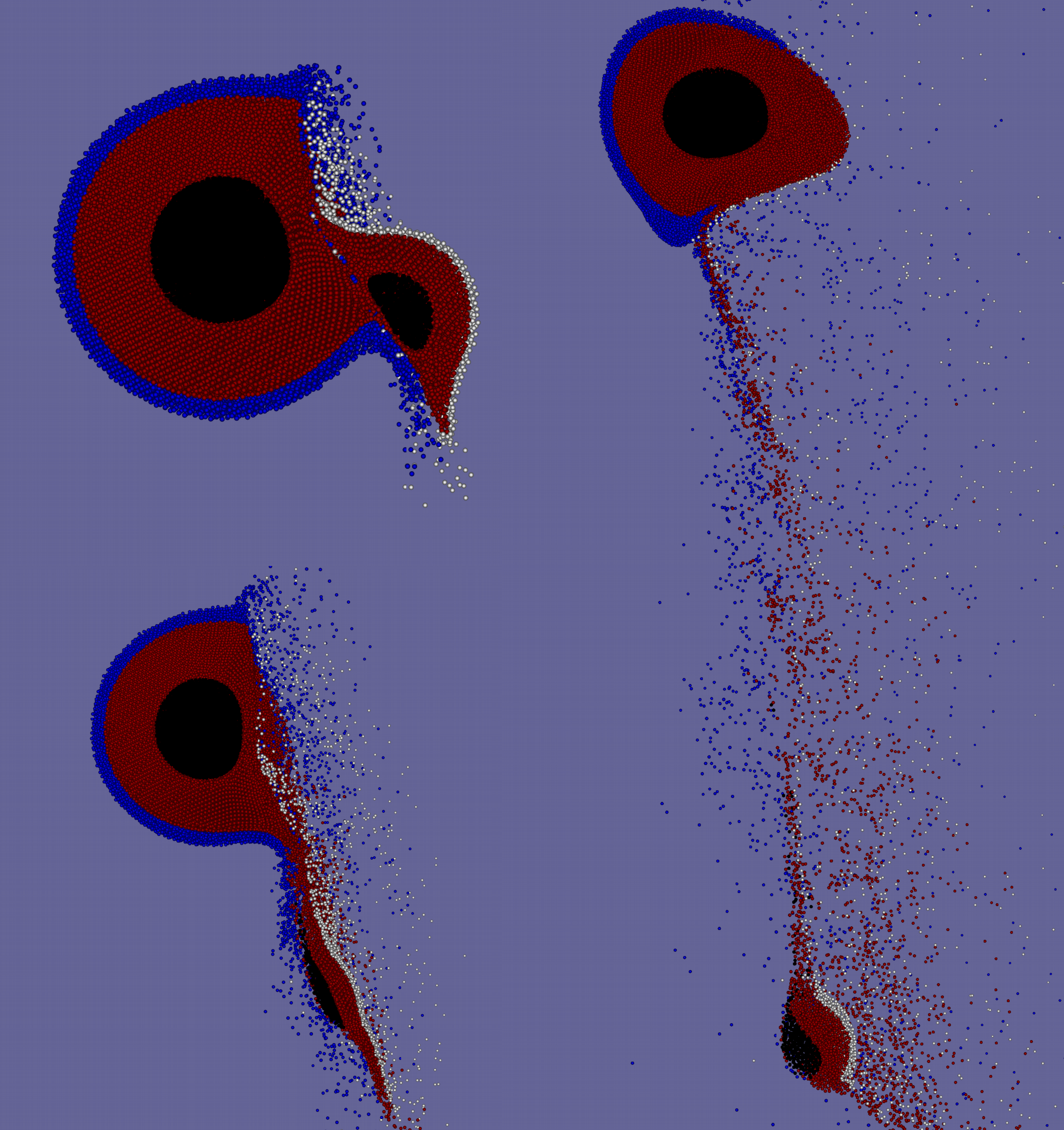}
	\includegraphics[width=0.34\hsize]{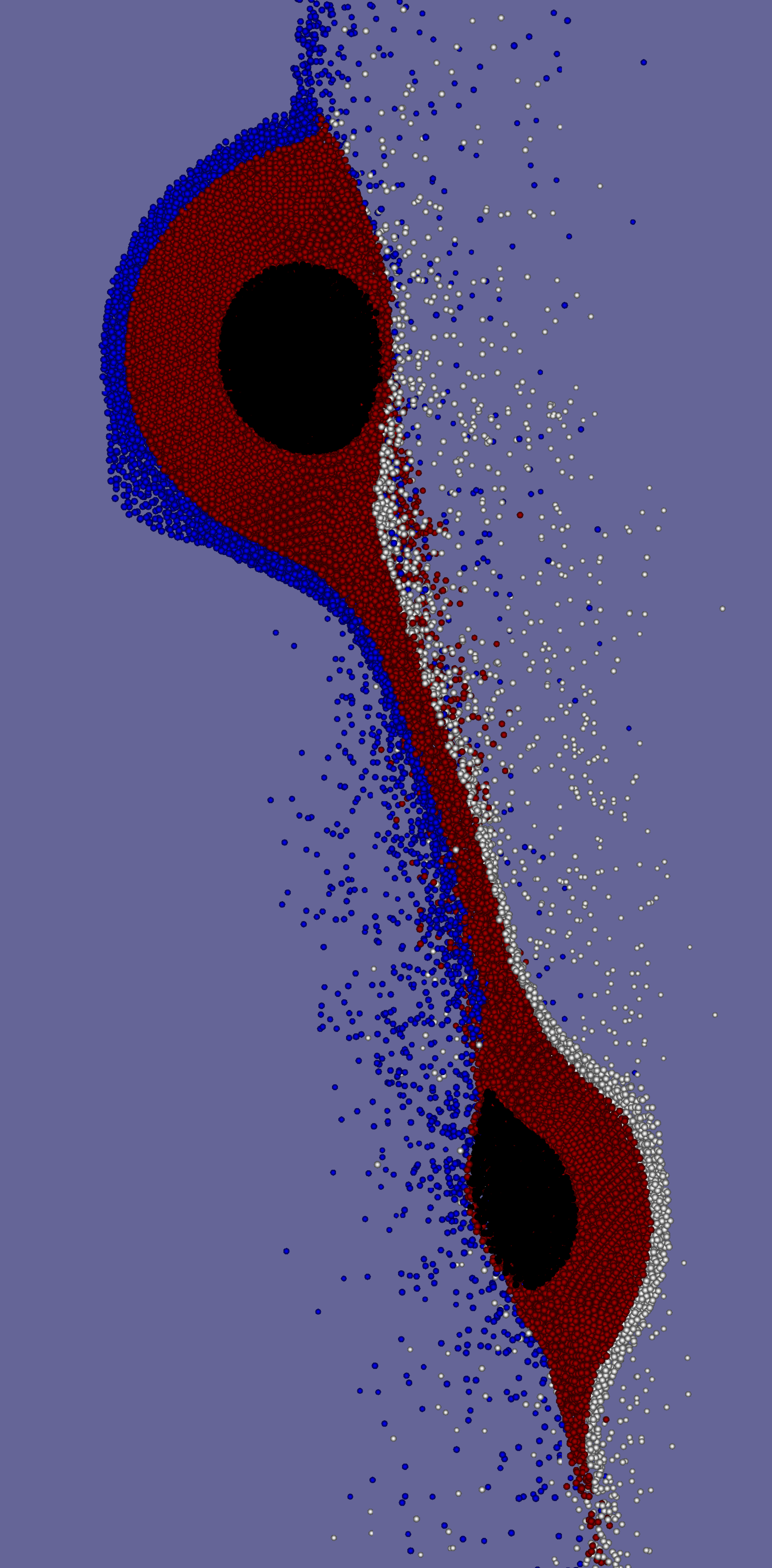}
	\caption{Simulation snapshots (cut views) to illustrate the mechanism of volatile transfer and loss. Water originally on the target/projectile in blue/white, and red and black particles represent basalt and iron, respectively (see the online version for color). The left sequence shows a scenario with $v/v_\mathrm{esc} = 2.5$, $\alpha = 45^\circ$ and $\gamma$ = 1:9, and the right image represents the same parameters except for $\gamma$ = 1:2 (the same scenarios as in the two lower panels in Fig.~\ref{fig:snapshots}). See the text for discussion.}
	\label{fig:cut_snapshots}
	\end{figure}

To explore the more subtle consequences of hit-and-run, like to what proportions a post-collision fragment's water inventory originated from projectile and target, we find it instructive to leave the target-projectile point-of-view and rather switch to a \emph{one-body perspective}.
In Fig.~\ref{fig:snapshots}, which illustrates the effects of varying the main collision parameters, projectile and target water are highlighted in different colors (see online version). This visualizes compositional mixing (transfer) of projectile and target water (see also Fig.~\ref{fig:cut_snapshots}), as well as losses from the individual pre-collision inventories.
Figure~\ref{fig:one_body_perspective} illustrates the consequences for a single body when hit by a range of different impactors, smaller and larger ones.
We denote the body of interest always as \emph{target} and the impacting one as \emph{projectile}, irrespective of which one is larger, and keep the definition $\gamma = M_\mathrm{proj}/M_\mathrm{targ}$.
The six panels in Fig.~\ref{fig:one_body_perspective} correspond to the six ($v/v_\mathrm{esc}$, $\alpha$) parameter pairs enframed in Fig.~\ref{fig:fancy_summary_m1e23}, and exemplify the effects on a single body for common hit-and-run parameters.
From the three graphs in each plot the topmost one shows the post-collision mass and water content if both bodies initially have a wmf of 0.1. Not surprisingly water losses are small for $\gamma < 1$, and strongly increase up to around 50\% for larger projectiles $\gamma > 1$, except for the calmest scenario (lower left panel).
The transition from $\gamma < 1$ to $\gamma > 1$ is smooth and no rapid increase in water losses -- as one might expect -- is found.
The middle graphs in Fig.~\ref{fig:one_body_perspective} represent the same collisions but accounting only for water initially on the target, hence they represent the impact of an entirely dry projectile onto a 0.1 wmf target.
The relatively small differences compared to the topmost graphs in most cases indicate only little transfer of water from projectile to target, even for large $\gamma \gg 1$, where the projectile could in principle provide large amounts of water to the target due to its size.
This is further emphasized in the bottom graphs, which correspondingly show the outcome if the target were initially dry.
Interestingly the target's wmf (originating solely in projectile water) increases with $\gamma$, peaks at relatively small, positive values of $\gamma$ and decreases again for even larger projectiles (with even larger volatile contents).
Except for rather central $\alpha = 30^\circ$ impacts the highest transferred water content is only around 1/10 of the (much larger) impactor's wmf.
In the $\alpha = 30^\circ$ collisions, especially in the low-velocity $v/v_\mathrm{esc} = 1.5$ case, there is considerably more transfer of volatiles, and towards $\gamma$ = 2:1 transfer from the projectile to the target becomes efficient and the combined wmf seems to even increase again.
However, obviously this situation occurs only for the lowest-velocity hit-and-run encounters, and indeed this scenario is on the edge to the graze-and-merge regime \citep{Leinhardt2012_Collisions_between_gravity-dominated_bodies_I}, indicated by post-collision velocities barely above $v_\mathrm{esc}$.
For all larger mass ratios, and thus lower collision energies, the outcome is not hit-and-run anymore (and therefore not plotted).
Hence it seems that transfer of volatiles between similarly-composed bodies in hit-and-run events has a rather minor influence compared to collisional erosion.

\subsection{Dependence on total mass}
While the bulk of this study focuses on collisions of $\sim$\,Moon-sized embryos ($M_\mathrm{tot} = 10^{23}$\,kg) we also considered different masses between $10^{22}$ and $10^{25}$\,kg, to study the influence of the total mass on the general collision outcome and especially on vaporization of volatile material (see~Sect.~\ref{sect:water_vapor}).
We already investigated the dependence of similar-sized collision outcomes on total mass in \citet{Burger2017_hydrodynamic_scaling} in detail, and will therefore limit ourselves to some important remarks in this paper.
The results presented in Fig.~\ref{fig:vaporization} indicate that for the larger body the outcome is very similar over the whole investigated range of masses -- their volatile inventory remains largely untouched (upper panels).
The situation for the smaller body is substantially different (lower panels).
For the three simulated mass ratios final wmf are only in the $\gamma$ = 1:2 case relatively independent of total mass, but show large variations for $\gamma$ = 1:9, ranging from 0.75 down to only 0.27, and for $\gamma$ = 1:20 even from 0.79 down to basically zero.
\changed{The main reasons for these differences are the transition from sub-sonic to super-sonic collision velocities with increasing mass (for other parameters equal), and increasing gravitational compression towards higher masses, resulting in more compact objects with greater hydrostatic pressures to be partly released upon impact.}
This clearly shows that the total mass is a crucial parameter for stripping of volatiles from the smaller body in a hit-and-run encounter, and at least this aspect of similar-sized collisions can certainly not be assumed to be scale-invariant \citep{Asphaug2010_Similar_sized_collisions_diversity_of_planets, Burger2017_hydrodynamic_scaling}.

\subsection{Dependence on water amount and distribution}
\label{sect:water_amount_distribution}
In order to check the influence of the chosen composition model, we performed additional simulation runs with entirely dry target bodies (with $v/v_\mathrm{esc} = 2.5$, $\gamma$ = 1:9, $M_\mathrm{tot} = 10^{23}$\,kg and $\alpha = 45^\circ$ and $60^\circ$; cf.~Tab.~\ref{tab:hit-and-run_results}), and compared them to runs with otherwise equal parameters and our usual composition model with both bodies covered in water (ice).
Naturally the overall post-collision water contents are lower now, but the amount transferred from the projectile to the target and the amount lost from the projectile are expected to be approximately equal, and independent of the target's precise composition.
Our results confirm this, with differences in wmf (and also in fragment masses and kinetics) around 20\% or lower. A dry target consistently accretes more water (from the projectile) than a water-rich one, and causes the projectile to loose a larger fraction of its initial volatile content. The projectile also looses more mass in general when colliding with a dry target instead of a water-rich one, where the difference is larger for the $\alpha = 45^\circ$ case than for $60^\circ$.
We suspect that this behavior is an interplay between the higher density of basalt compared to water, which results in a larger resistance against impacting material, and a slightly different collision geometry, because a dry target is more compact than a volatile-rich one. The latter means not least a higher specific impact energy in the dry-target runs (e.g. in the $\alpha = 45^\circ$ case about 5\%, 1.27 vs.\ 1.33\,MJ/kg), since all other parameters were kept constant. This is consistent with the higher losses in the dry-target run.

A point closely related to the above is the influence of the extent of the water layer. We repeated the $v/v_\mathrm{esc} = 2.5$, $\alpha = 45^\circ$, $\gamma$ = 1:2, $M_\mathrm{tot} = 10^{23}$\,kg scenario with wmf of 5\% and 20\%, in addition to the standard 10\% run (see Tab.~\ref{tab:hit-and-run_results}), and found only relatively small deviations from the expectation that relative water losses and transfers are similar. This means for instance that the post-collision water content of the target is roughly twice as large for the 10\% water scenario than for the 5\% one, and again approximately doubled for the 20\% run.
Considering all three scenarios, deviations are around 25\% at most, but if only the 5\% and 10\% water runs are compared the differences reduce to 10\% and less, suggesting a probable convergence towards increasingly thinner volatile layers.
The general trend and the expected underlying reasons are the same as in the dry-target case above. Bodies with thinner water ice shell accrete more (from the other body), but also loose more water from their initial inventory, probably again due to slightly different specific impact energies (2.96 vs.\ 2.86 vs.\ 2.70\,MJ/kg).
\citet{MaindlSchaeferHaghighipourBurger2017_water_transport} also studied the influence of different initial volatile contents and also found only small variations ($\sim$\,10\%) in relative water losses, in concordance with our results.
These authors additionally considered water SPH particles randomly distributed inside the whole projectile body.
We did not include such uniform distributions because there is strong evidence that bodies in the mass range considered here are probably largely differentiated. It is also yet unclear how such water inclusions (e.g.\ in hydrated minerals) can be accurately modeled, and how related and perhaps important effects like degassing due to pressure unloading can be included in a consistent way \citep[but see e.g.][]{Asphaug2006_hit-and-run}.
However, at least for collisions in the hit-and-run regime \citet{MaindlSchaeferHaghighipourBurger2017_water_transport} found no large differences between randomly distributed water and spherical shell configurations in terms of water losses.
The generally low dependence of final (relative) wmf on the extent and distribution of volatile inventories is an important result, which helps to safely reduce the dimensionality of the parameter space in simulations on water delivery. We discuss this further in Sect.~\ref{sect:discussion_conclusions}.

\subsection{Influence of material strength}
\label{sect:material_strength}
The justification for usually modeling sufficiently large bodies as strengthless fluids is motivated by highly dominating gravitational stresses over material strength. \citet{Jutzi2015_pressure_dependent_failure_models} studied collisions between similar-sized bodies and found that including a realistic rheology is still necessary for 100\,km bodies, and recent results using the same SPH code and material model \citep{Burger2017_hydrodynamic_scaling} have shown significant differences in fragment characteristics and also in water losses between solid-body simulations and (otherwise identical) purely hydrodynamic runs also for embryo-sized bodies.
For a more comprehensive overview we refer to the latter study, and discuss the topic only briefly here, exemplified by two scenarios (see~Tab.~\ref{tab:hit-and-run_results}) which were computed again with the full solid-body model as outlined in Sect.~\ref{sect:methods}.
The results are generally in good agreement with their strengthless counterparts, and differ by less than 20\% for final wmf and losses, which is also consistent with the results of \citet{Burger2017_hydrodynamic_scaling}.
Larger deviations of up to 50\% are found only for the contribution of transferred volatiles (between projectile and target) to the final inventory.
However, this is a more subtle process, and only a minor contribution to a fragment's post-collision water budget compared to the retained amount (cf.~Fig.~\ref{fig:one_body_perspective}), therefore these relatively large variations are not surprising.
These contributions by transferred water are always smaller in the solid scenarios than in the respective strengthless runs, probably because tensile and shear strength act against removal and subsequent transfer.
It is yet unclear to what extent and for which aspects of giant collisions material strength becomes important. The large pressures in the interiors of sufficiently massive bodies change the behavior of geologic materials towards high viscosities and ductile, plastic flows \citep{Holsapple2009_strength_review}, and in general their rheology exhibits a complex dependence on stress, strain, strain-rate, temperature and pressure.
\changed{In addition even fully damaged \emph{rubble pile} material is often not strengthless, but can support considerable shear stresses.
This makes the here-used von Mises yield criterion not an ideal choice, as already indicated in Sect.~\ref{sect:methods}, since it does not consider pressure-dependent shear strength.
A more sophisticated model for the calculation of shear strength of geologic materials has been developed e.g.\ by \cite{Collins2004_modelling_damage_and_deformation}, who use a pressure-dependent yield strength for intact rock and a Coulomb dry-friction law for completely fragmented material.
We suggest and plan a dedicated future study to clarify the influence of different rheology models also beyond sizes of the 100\,km studied by \citep{Jutzi2015_pressure_dependent_failure_models}, including also volatile material like water ice.}
In addition to these complexities, the physical state of volatile inventories prior to the collision is also uncertain, where for instance a considerable fraction might be molten, which would put the application of material strength into perspective.
The behavior of real objects probably lies somewhere between that of a strengthless fluid and a fully solid body, thus these two models may be considered rather limiting cases, where further studies are necessary to clarify the necessity for certain material models.

\subsection{Water vapor production}
\label{sect:water_vapor}
Once impact energies rise above the vaporization energy of water (ice) large-scale vapor production can be expected.
\changed{Here we treat only vaporization caused directly by the impact and the further development of heat-redistribution, like melting or vaporization of ice by an impact-heated mantle, is not included.}
To exemplarily estimate the post-collision water vapor fraction we ran scenarios with fixed $v/v_{esc} = 2.5$ and $\alpha = 45^\circ$ for several different total masses (and thus impact energies) between $\sim\,\!M_\mathrm{Moon}/10$ and $M_\mathrm{Earth}$ and for varying mass ratios of 1:2, 1:9 and 1:20 (see~Tab.~\ref{tab:hit-and-run_results}).
	\begin{figure}
	\centering
	\includegraphics[width=\hsize]{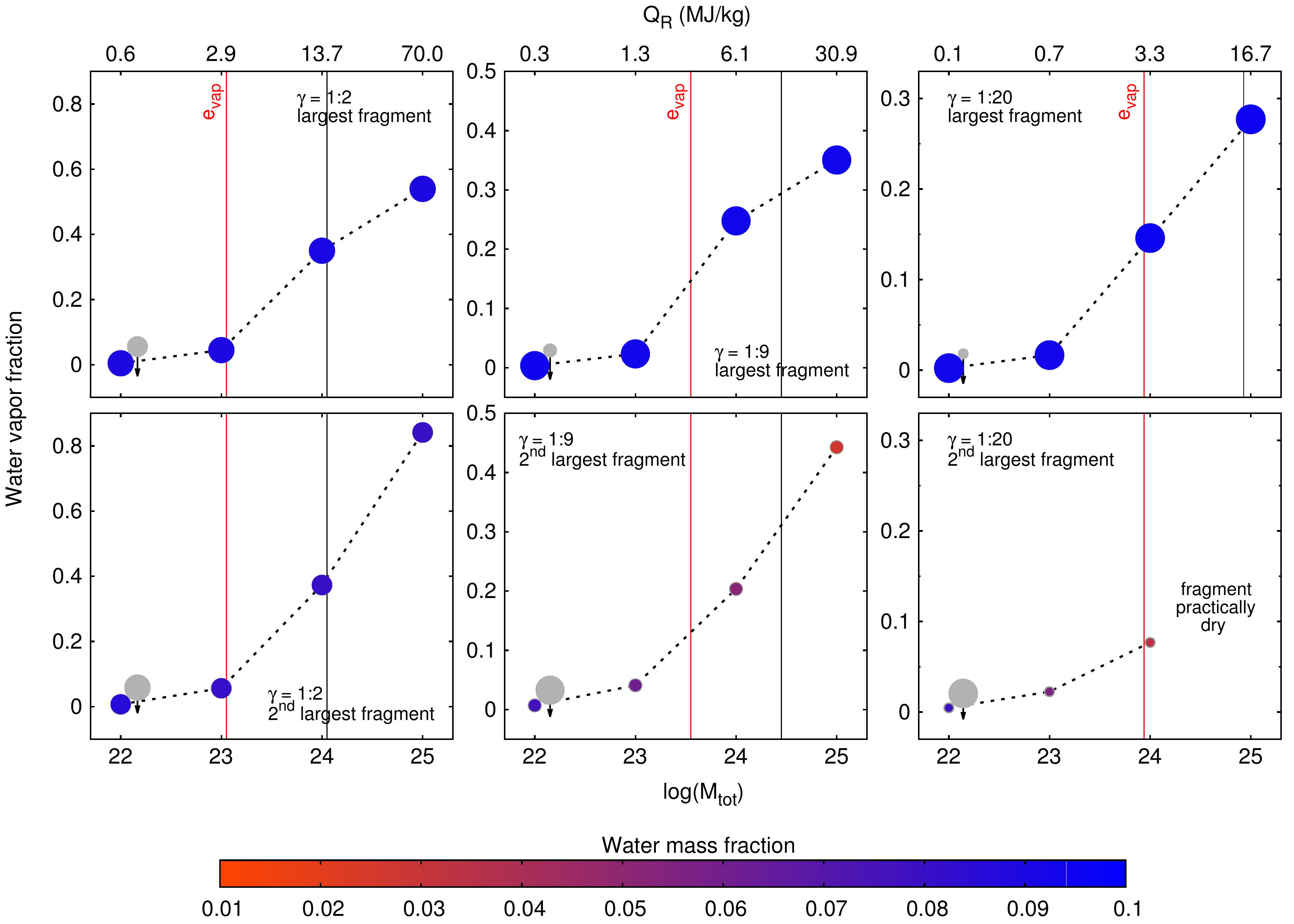}
	\caption{Collisionally produced fraction of water vapor in the two largest fragments (relative to their total water content), plotted over $M_\mathrm{tot}$ for three different mass ratios (columns). Impact velocity and angle are fixed at $v/v_\mathrm{esc} = 2.5$ and $\alpha = 45^\circ$, as depicted by the impacting projectile as light gray circles. The specific collision energy $Q_R$, its special value $e_\mathrm{vap} = e_\mathrm{cv}$ (see the Tillotson eos in Sect.~\ref{sect:methods}), and $Q_R = 15$\,MJ/kg are also indicated by vertical lines. See the online version for color.}
	\label{fig:vaporization}
	\end{figure}
The results are summarized in Fig.~\ref{fig:vaporization}, where the reduced mass specific impact energy $Q_R$ of the individual scenarios is also indicated. It is given by \citep{Stewart2009_Velocity-dependent_catastrophic_disruption_criteria}
\begin{equation}
\label{eq:Q_R}
Q_R = \frac{\mu\, v_0^2}{2\, M_\mathrm{tot}}\ ,
\end{equation}
with the reduced mass $\mu = M_\mathrm{proj} M_\mathrm{targ}/M_\mathrm{tot}$ and collision velocity $v_0$ (see Fig.~\ref{fig:collision_geometry}).
Not surprisingly the water vapor fraction strongly increases with $M_\mathrm{tot}$, and larger amounts of water vapor are only produced for energies greater than $e_\mathrm{vap}$. At this threshold around 10\% of the water inventories are already vaporized on both large fragments, and roughly equal throughout the three investigated mass ratios (note the different y-axis scales).
If considered as a function of $M_\mathrm{tot}$ the water vapor content is a function of $\gamma$ as well, because $Q_R$ is also a function of $\gamma$ (for all other parameters equal), as elaborated in Sect.~\ref{sect:water_transfer_and_loss} (see footnote~\ref{footnote:Q_R_of_gamma}), and there is certainly a strong correlation between the available energy and the amount of vaporization.
The results in Fig.~\ref{fig:vaporization} indicate that water vapor production seems to scale indeed well with $Q_R$, relatively independent of the mass ratio, well visible when comparing vapor fractions for specific $Q_R$ values (like indicated by the vertical lines).
Also the individual vapor fractions on the two largest fragments are surprisingly similar, also for smaller mass ratios.
From the remaining water on the two large fragments after a hit-and-run roughly up to 40\% can be vaporized for Moon to Mars-sized embryos, and up to 80\% in collisions involving Earth-mass objects. These numbers, however, depend on the mass ratio (for a fixed total mass) and may also deviate significantly for other values of impact velocity and angle.
\changed{Note that due to the limitations of the Tillotson eos the computed vapor fractions are rather rough estimates, as a first step indicating possible directions for future work.}

\subsection{Dependence on resolution}
\label{sect:resolution}
\changed{
	\begin{figure}
	\centering
	\includegraphics[width=1.0\hsize]{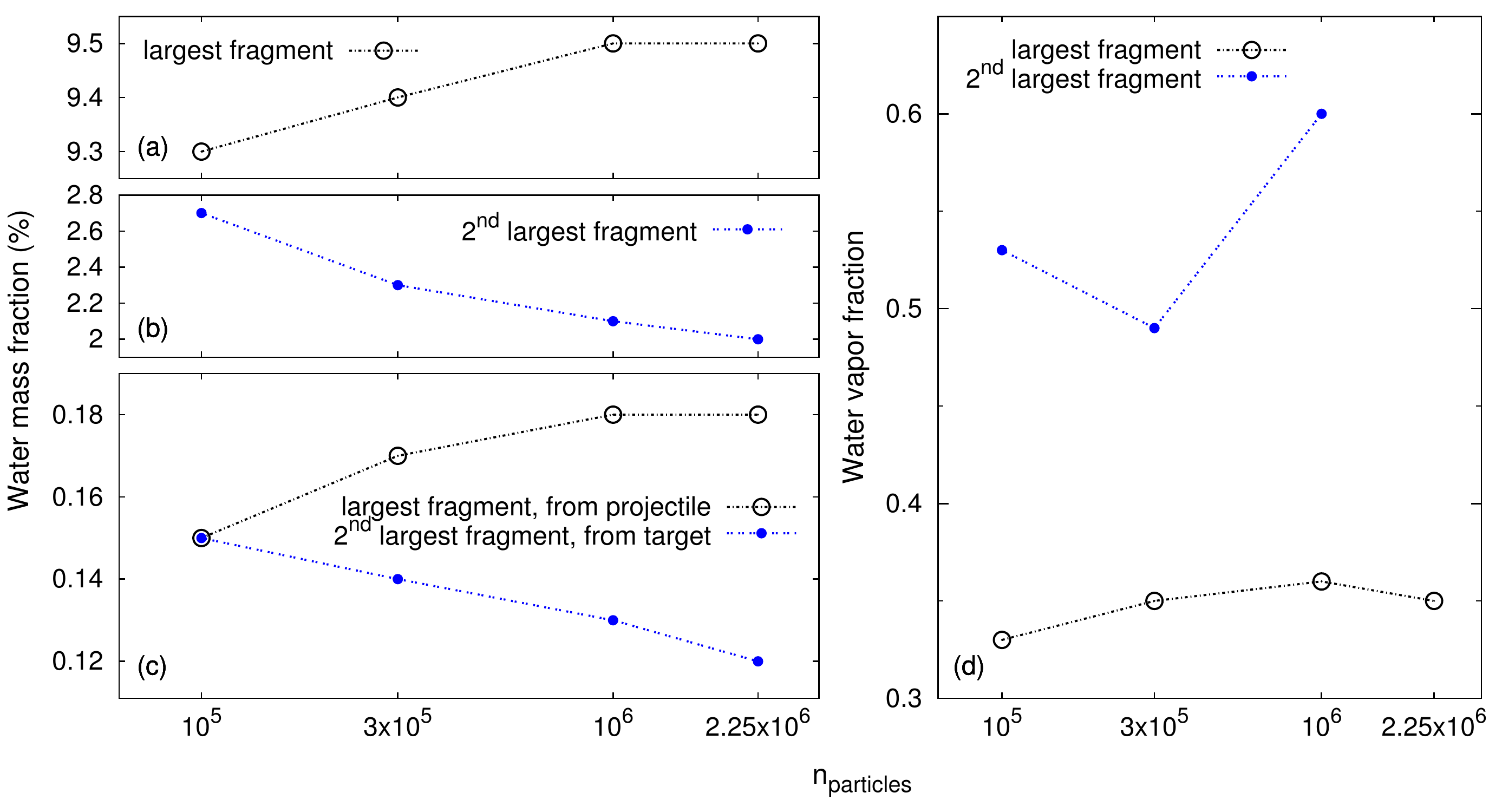}
	\caption{Dependence on resolution for the outcome of the scenario $M_\mathrm{tot} = 10^{25}$\,kg, $\alpha = 45^\circ$, $v/v_\mathrm{esc} = 2.5$ and $\gamma$ = 1:9 (see Tab.~\ref{tab:hit-and-run_results}). Panels (a) and (b) show the wmf of the two largest fragments, panel (c) the transferred wmf originating only from the other body (see Sect.~\ref{sect:water_transfer_and_loss} and \ref{sect:water_amount_distribution}), and (d) water vapor fractions (see Sect.~\ref{sect:water_vapor}) of the two largest fragments after the collision.}
	\label{fig:resolution_plots}
	\end{figure}
The resolution in SPH simulations is defined by the particle number. 
\citet{Genda2015_Resolution_dependence_of_collisions} investigated the influence of resolution on the critical specific impact energy for disruption (where the largest post-collision fragment has half the total mass) for gravity-dominated bodies, and found a factor of two difference between 50k and 5\,million SPH particles, due to different efficiencies of kinetic energy dissipation. They conclude that approximate convergence is reached only for their highest resolutions.
Especially low-density regions, like impact ejecta, are affected by differing particle numbers, where the evolution in such regions is often dominated by resolution instead of physics \citep{Reinhardt2017_Numerical_aspects_of_giant_impact,Genda2015_Resolution_dependence_of_collisions}.
Another closely connected issue is correctly resolving shock waves, where about 10 particles are required in one dimension \citep[e.g.][]{Genda2015_Resolution_dependence_of_collisions}.
While our standard resolution (100k) is sufficient for projectile and target diameters being always clearly above this threshold, the thickness of the water envelope alone is usually below it.
For example a single body made of $n_\mathrm{part}=10^5$ particles and a wmf of 0.1 has a diameter $\propto n_\mathrm{part}^{1/3}$ of $\sim$\,57 particles, but a water shell thickness of only $\sim$\,4 particles, while the latter figure is well above 10 for our highest resolution of 2.25\,million.

To check the reliability of our results we ran three selected scenarios (see~Tab.~\ref{tab:hit-and-run_results}) also with increased resolutions of 300k, 1\,million and one with 2.25\,million particles (in addition to the standard 100k runs).
The outcome for several main quantities of interest as a function of particle number is plotted in Fig.~\ref{fig:resolution_plots} for a collision between an $\sim$\,Earth-sized and a $\sim$\,Mars-sized object, indicating approximate resolution convergence for most quantities. The large masses of the involved bodies ($M_\mathrm{tot} = 10^{25}$\,kg) results in a very energetic collision and thus in large-scale water vapor production (Sect.~\ref{sect:water_vapor}), ideal for clarifying the dependence on resolution for the purposes of our topic.
In all three scenarios the global outcomes (fragment masses and their kinetics) of runs with different resolutions show only minor deviations below 10\%. Post-collision wmf of the largest fragment are even more accurate within only few \%, and for the second-largest fragment and the wmf transferred between projectile and target results are still within $\sim$\,25\% (cf.\ Fig.~\ref{fig:resolution_plots}).
Note that the outcomes of the 100k runs seem to give an upper limit for the second-largest fragment's wmf, meaning that resolution converged volatile stripping of the smaller of the colliding bodies is even more efficient than indicated by the 100k results.
The situation for the water vapor fraction is twofold, with deviations of less than 10\% for the largest fragment, and $\sim$\,20\% for the second-largest one.
A more detailed analysis of the latter showed that these values take particularly long to converge after the actual collision itself, since ongoing accretion of debris continues to dissipate energy, leading to further vaporization even after the two main fragments have clearly separated and the material bridge between them has largely dissolved (cf.\ Fig.~\ref{fig:snapshots}). For the 2.25\,million particles run this quantity has still not converged at the end of the simulation time and is therefore not included in Fig.~\ref{fig:resolution_plots}.
We conclude that simulations with probably even higher resolutions would have to be ran for particularly long (simulated) times to finally clarify this issue.

The deviations for the different quantities described above can also be considered a rough error estimate of the presented results w.r.t.\ resolution convergence.
The results confirm that shock acceleration and energy dissipation are already fairly well resolved in the 100k runs, also for the relatively thin (in terms of particle layers) water shells at this resolution. Albeit true resolution convergence is hard to proof these tests indicate that our results are in general within $\sim$\,25\% of this limit.}

\subsection{Comparison with collision outcome models}
\label{sect:outcome_model}
The most basic collision outcome model is perfect merging with conservation of linear momentum, and the assumption of 100\% retention of volatiles. This is certainly not realistic in most situations.
\citet{Leinhardt2012_Collisions_between_gravity-dominated_bodies_I} and \citet{Stewart2012_Collisions_between_gravity-dominated_bodies_II} developed a comprehensive analytical model that distinguishes several collision outcome regimes, including partial accretion, erosion and hit-and-run. It has been included in N-body simulations and used in several studies on planet formation \citep[e.g.][]{Dwyer2015_Bulk_chemical_consequences_of_incomplete_accretion, Bonsor2015_Collisional_origin_to_Earths_non-chondritic_composition, Leinhardt2015_Numerically_predicted_signatures_of_TPF}. It is based on scaling laws to determine the mass of the largest collisional fragment $M_\mathrm{lf}$, and also applies corrections for varying mass ratios as well as the reduced interacting mass in oblique collisions. To also track basic changes in bulk composition they suggest to include a simple model for mantle stripping introduced by \citet{Marcus2010_Icy_Super_Earths_max_water_content}, which considers two idealized cases for the core mass: (1) $M_\mathrm{core} = \mathrm{min}(M_\mathrm{lf}, M_\mathrm{core,proj}+M_\mathrm{core,targ})$, i.e.\ mantle is only added to the largest fragment once both cores are used up, and (2) either $M_\mathrm{core} = M_\mathrm{core,targ} + \mathrm{min}(M_\mathrm{core,proj}, M_\mathrm{lf} - M_\mathrm{targ})$ on accretion (i.e.\ if $M_\mathrm{lf} > M_\mathrm{targ}$), or $M_\mathrm{core} = \mathrm{min}(M_\mathrm{core,targ}, M_\mathrm{lf})$ for target erosion (i.e.\ if $M_\mathrm{lf} < M_\mathrm{targ}$). This means model (2) first adds projectile core material to the whole target in case of accretion, but removes material from the target (starting with the mantle) for erosion.
To compute the actual change in composition it is suggested to use the average of (1) and (2).
For the second large survivor in hit-and-run events they suggest to consider the reverse impact, where the projectile is hit by a hypothetical body consisting only of the part of the target that geometrically overlaps with the projectile, and to apply the above model to this (reversed) collision situation to determine compositional changes of the second largest fragment.
Even though this model was not directly developed for volatile transfer and loss in hit-and-run collisions, but rather for the less subtle effects of major bulk compositional changes, we compare it to our numerical results.
It turned out that it gives only a very crude estimation of the changes in water contents for scenarios like ours. While predictions for the target body are mostly at least in broad agreement with numerical results, those for the projectile are often very far off. Therefore we have not included these predictions in the one-body perspective results (Fig.~\ref{fig:one_body_perspective}), but only in the combined view in Fig.~\ref{fig:water_loss_over_vesc}, as disconnected and smaller but otherwise equal symbols for comparison.
The predicted combined water losses are mostly much larger than the numerical results, mainly because the reverse impact is predicted much more destructive than it actually is, leaving the second largest fragment often entirely water-stripped in this model. We suspect that this is at least partly due to the geometry of a hit-and-run event, where the interacting fraction of the impacting body still grazes by the impacted one, which is likely to be less destructive than a more central collision with the same impacting mass.
In addition this model is based on absolute masses of the cores (which is everything except the water layer for our purposes) and the (predicted) largest fragment, thus uncertainties increase the lower the amount of the material considered for stripping is.
Also, taking a closer look at the components (1) and (2) and the analysis and impression of hit-and-run collisions (cf.~Fig.~\ref{fig:cut_snapshots}), it seems that outcomes are typically rather close or even \emph{beyond} predictions of component (2) instead of an average of (1) and (2), for both accretion of projectile material by the target and also for target erosion.
Note also that the model can not be reasonably used for volatile transfer, for instance from a water-rich projectile to a dry target.

\section{Discussion and conclusions}
\label{sect:discussion_conclusions}

\subsection{Volatile transfer and loss}
Our results show that volatile loss and transfer in hit-and-run collisions is a function of four main parameters, the impact velocity, angle, the colliding bodies' mass ratio, and also of the total colliding mass.
The results for combined water loss in Fig.~\ref{fig:water_loss_over_vesc} clearly illustrate that losses increase with impact velocity and mass ratio, and decrease with impact angle. 
We find significant values in most parts of the parameter space, except for the most grazing impacts ($\alpha \gtrsim 60^\circ$), lowest mass ratios ($\lesssim$ 1:50) and for impact velocities close to $v_\mathrm{esc}$.
\changed{In a common protoplanet encounter \citep[e.g.][]{Maindl2014_Collision_parameters} with $\alpha = 45^\circ$, $\gamma$ = 1:9 and $v/v_\mathrm{esc}$ between 1.5 and 2.5, the water mass fraction (wmf) of the target decreases only by less than $\sim$\,10\%, while the projectile looses between 5\% and up to 40 - 50\% of its initial wmf in this velocity range (Fig.~\ref{fig:one_body_perspective}). However, it is important to keep in mind that this figure can change over a wide range (up as well as down) for sufficiently different scenarios.}
Figure~\ref{fig:vaporization} shows that losses also increase strongly with the colliding bodies' mass, which is mainly due to strong shocks once impact speeds become supersonic. For our scenarios the speed of sound is around 3\,km/s. For $v/v_\mathrm{esc} = 2.5$, impact velocities approach this value for $M_\mathrm{tot}$ between $10^{22}$ and $10^{23}$\,kg.
Our results for overall water loss are in good agreement with previous studies.
The reason why head-on and also low-obliquity hit-and-run impacts strip the most volatiles is particularly associated with the large interacting mass in such events compared to more grazing encounters. The cut views in Fig.~\ref{fig:cut_snapshots} visualize this for a $\gamma$ = 1:9 and 1:2 scenario.
The smaller the projectile compared to the target the more of it is directly affected and mechanically disrupted due to enormous shear stresses. Additionally shocks and gravitational (tidal) stresses enhance disruption. For $\gamma$ = 1:9 in Fig.~\ref{fig:cut_snapshots} (left) the projectile is ripped apart down to its core, while for $\gamma$ = 1:2 and the same $\alpha$ (right) both bodies stay more or less intact.
During the main interaction phase the projectile ploughs through the target (and the target through the projectile) and both shear away the outer layers of their opponent and accelerate most of it away, while also dragging some material with them. This is how the bulk of the volatiles accreted from the other body are transferred.
It is well visible from Fig.~\ref{fig:cut_snapshots} that the water shell is rather unaffected where no direct mechanical interaction with the impactor took place (even though strong shocks can blow parts of this off too).
It is due to this behavior that approximately equal fractions of the overall volatile inventory are lost or transferred even for different extents of the water shell (cf.~Sect.~\ref{sect:water_amount_distribution}).

\citet{Marcus2010_Icy_Super_Earths_max_water_content} found that volatile fractions practically never increase in giant collisions of similarly composed Earth to Super-Earth-mass bodies, which we can confirm for all our scenarios, and this also holds for the two large fragments in a hit-and-run individually. Note that this does not hold anymore if pre-collision wmf are sufficiently different. The simulations with dry targets (Sect.~\ref{sect:water_amount_distribution}) are an example thereof.
While embryo-sized and larger objects become increasingly difficult to disrupt by smaller impactors \citep{Asphaug2006_hit-and-run}, which we can confirm also for their volatile inventories, hit-and-run collisions are highly transformative for the smaller one of the colliding pair \citep[e.g.][]{Asphaug2010_Similar_sized_collisions_diversity_of_planets}.
We find that this holds especially also for stripping of outer volatile layers, like the water shells in our scenarios.
The inefficiency of transfer of projectile volatiles to the target as long as pre-collision mass fractions are similar (illustrated in Fig.~\ref{fig:one_body_perspective}), means that hit-and-run post-collision inventories in these cases are dominated by the bodies' pre-collision inventories reduced by impact losses. These impact losses are an increasing function of mass ratio for the whole range of simulated values between 1:50 and 50:1.
Interestingly the transferred wmf on the contrary exhibits a maximum value for a given body but differently massive projectiles, and decreases again for even larger impactors despite their greater (absolute) volatile inventory. This peak is reached for projectiles a couple of times more massive than the target (Fig.~\ref{fig:one_body_perspective}), and even in these cases the target can accrete a wmf of only up to 10\% of the projectile's pre-collision value, except for slow, low-obliquity encounters.
However, transferred volatiles can still significantly modify isotopic fingerprints, while deeper, more protected regions are not or only little affected.

\subsection{Vaporization of volatile inventories}
Losses due to escape of collisionally vaporized material have not been included so far in studies on volatile loss in giant collisions.
Our results indicate that the produced vapor fraction is similar on both large fragments, even for smaller mass ratios (Fig.~\ref{fig:vaporization}).
To assess how much of this vaporized material eventually escapes, remains in the object's atmosphere or perhaps recondenses, a lot of additional factors come into play, among them the mass of body and atmosphere, their thermal state \changed{(e.g.\ further vaporization by an impact-heated mantle)}, and external factors as the distance to the host star, the stellar EUV flux and the frequency of impacts.
Recent modeling by \citet{Odert2017_Escape_fractionation_of_volatiles} that takes most of these factors into account suggests that embryos in the terrestrial planet region up to Mars-size quickly loose their water vapor atmosphere ($\sim$\,Myrs), either thermally or driven by the large EUV fluxes of young stars.
Due to the large number of unknowns in this figure and the only approximate treatment of vaporization with the Tillotson eos, we can not make quantitative conclusions on these additional volatile losses.
However, our results (Fig.~\ref{fig:vaporization}) indicate that vaporization losses are probably most important in our intermediate mass scenarios around $M_\mathrm{tot} = 10^{24}$\,kg ($\sim$ Mars-sized bodies), where vapor fractions are already considerable, but the (hit-and-run) fragments' masses, particularly of the smaller body, are still low enough for large scale escape.
According to our estimates fragments in this mass range could loose another $\sim$\,10 - 40\% of their remaining volatile inventory.
Therefore it seems that collisional volatile stripping is dominated by direct losses due to mechanical/gravitational stresses and shock acceleration, but vaporization and subsequent loss can still enhance this further, not as the dominating contribution, but also not negligibly.
\changed{Continuing from these first steps, a dedicated future study including an improved treatment of phase changes by a thermodynamically consistent eos like ANEOS \citep[see e.g.][]{Melosh2007_A_hydrocode_EOS_for_SiO2} and higher resolution to resolve also low-density plumes in detail \citep{Reinhardt2017_Numerical_aspects_of_giant_impact} could provide improved estimates on this topic.}

\subsection{Water delivery to terrestrial planets}
What does this mean for water (volatile) delivery in early planetary systems?
With a focus on the \emph{target planet} for water delivery the prime example is proto-Earth, with strong isotopic evidence that the bulk of its water inventory originated from the outer main belt (or the same source as this material).
In a scenario where the majority of these volatiles is delivered by only a relatively small number of large impactors \citep{Morbidelli2000_Source_regions_for_water}, hit-and-run can play a decisive role, because it is a frequent outcome in such similar-sized collisions \citep[$\sim$\,50\%;][]{Kokubo2010_Formation_under_realistic_accretion,Genda2017_Hybrid_code_Ejection_of_iron-bearing_fragments}.
The chances for growing planets like proto-Earth to be hit by a larger body are slim, therefore the majority of impactors are likely smaller, and the generally low transfer efficiency, combined with additional losses as from vaporization, strongly limits the amount of accreted volatiles.
This implies a substantial difference compared to the still mostly used assumption of perfect merging of water inventories in (N-body) planet formation simulations.
It depends on the dynamical environment and resulting impact velocities, how efficient volatiles can be accreted from a hit-and-run impactor, but in such a scenario they might be high since the impactors were scattered over large orbital distances.
For collisions not in the hit-and-run regime the results for head-on scenarios in Fig.~\ref{fig:water_loss_over_vesc} indicate that volatile losses for rather central impacts are also high, but this is only the combined value (i.e.\ for identical pre-collision wmf of both bodies).
A more detailed study on the transfer efficiency in collisions beyond hit-and-run (less oblique, either accretionary or erosive) could help to reduce these uncertainties.
The consequences of hit-and-run for objects just below the top of the size distribution when colliding with the largest are often large-scale stripping of their volatiles. Our results confirm that this suggests a tendency towards dry Earth-mass planets in extrasolar systems with Super-Earths \citep[e.g.][]{Asphaug2010_Similar_sized_collisions_diversity_of_planets}.

In a young planetary system that is rather viewed as an ensemble of interacting embryos/protoplanets, the collisional evolution is also strongly driven by the distribution of impact velocities.
Even for modest values of $v/v_\mathrm{esc}$ combined water losses in hit-and-run events can go up to $\sim$\,40\% (Fig.~\ref{fig:water_loss_over_vesc}). 
It may not be unlikely that a specific embryo suffers multiple hit-and-run events, often as the smaller opponent, stripping most or all of its volatile content, before it is finally incorporated into a growing planet.
If bodies experience a series of such encounters, either as the larger or the smaller of the colliding pair, combined with additional losses of vaporized material and other environmental effects, it is probably a rather conservative estimate to place final water contents after the cumulative effects of a tens-of-Myrs giant collision phase at $\sim$\,1/10 of what perfect merging simulations suggest.
A combination of these loss processes is a natural explanation for the too high water contents that typically result from perfect merging simulations.
Currently available collision outcome models can not reliably predict volatile losses and changes in wmf, as shown by a comparison with our numerical results (Sect.~\ref{sect:outcome_model} and Fig.~\ref{fig:water_loss_over_vesc}), particularly not for the second largest hit-and-run fragments.
These models were developed rather for large-scale changes in bulk composition and not for the more subtle effects concerning (surface) volatile inventories.
\changed{Hybrid codes that simulate both, the long-term dynamical evolution as well as the hydrodynamics of individual collisions, are computationally very expensive and can thus be ran only in low-resolution currently, as recently done by \citet{Genda2017_Hybrid_code_Ejection_of_iron-bearing_fragments} with 10k SPH particles per collision. This is probably a too low resolution for studying the fate of surface volatiles.}
Realistic modeling of volatile delivery in planet formation simulations must begin with a profound understanding of transfer and loss mechanisms in individual collisions, to eventually include models developed for this particular aspect of collision outcomes into terrestrial planet formation computations.
\changed{The high dimensionality of the collision parameter space makes the development of these models -- e.g.\ in the form of scaling laws -- a formidable task, which we plan to address as the next logical step in a future study.}
The similarity in (relative) transfer and loss results (Sect.~\ref{sect:water_amount_distribution}), independent of the actual extent of the volatile shell (on either body), might proof very helpful for the development of such a model, because it likely eliminates the necessity to include absolute volatile amounts and their distribution as additional parameters.

\begin{acknowledgements}
We thank the anonymous reviewers for valuable comments and suggestions.
C.~B. and T.~I.~M. acknowledge support by the FWF Austrian Science Fund project S11603-N16.
The authors also acknowledge support by the High Performance and Cloud Computing Group at the Zentrum f{\"u}r Datenverarbeitung of the University of T{\"u}bingen, the state of Baden-W{\"u}rttemberg through bwHPC and the German Research Foundation (DFG) through grant no.\ INST 37/935-1 FUGG.
\end{acknowledgements}

\changed{
\section*{Appendix A -- Semi-analytical relaxation of initial conditions}
Simulations of giant collisions during the late stages of planet formation comprise large, self-gravitating bodies, which naturally exhibit a certain internal structure representing hydrostatic equilibrium. The initial conditions for such computations should be in a relaxed configuration (in terms of the applied numerical model) to resemble reality. 
The common approach is to apply some dynamical settling (numerical relaxation) prior to the actual simulation run, which, however, results in considerable amount of additional computing time.
To overcome the drawbacks of numerical relaxation we use a semi-analytical approach instead, by computing hydrostatic equilibrium structures for multi-material spherical bodies, before setting up the initial particle configuration according to them.
This is done in a self-consistent manner, i.e.\ consistent with the physical model implemented in the simulations themselves, to ensure a configuration already very close to equilibrium.
	\begin{figure}
	\centering{\includegraphics[width=1.0\columnwidth]{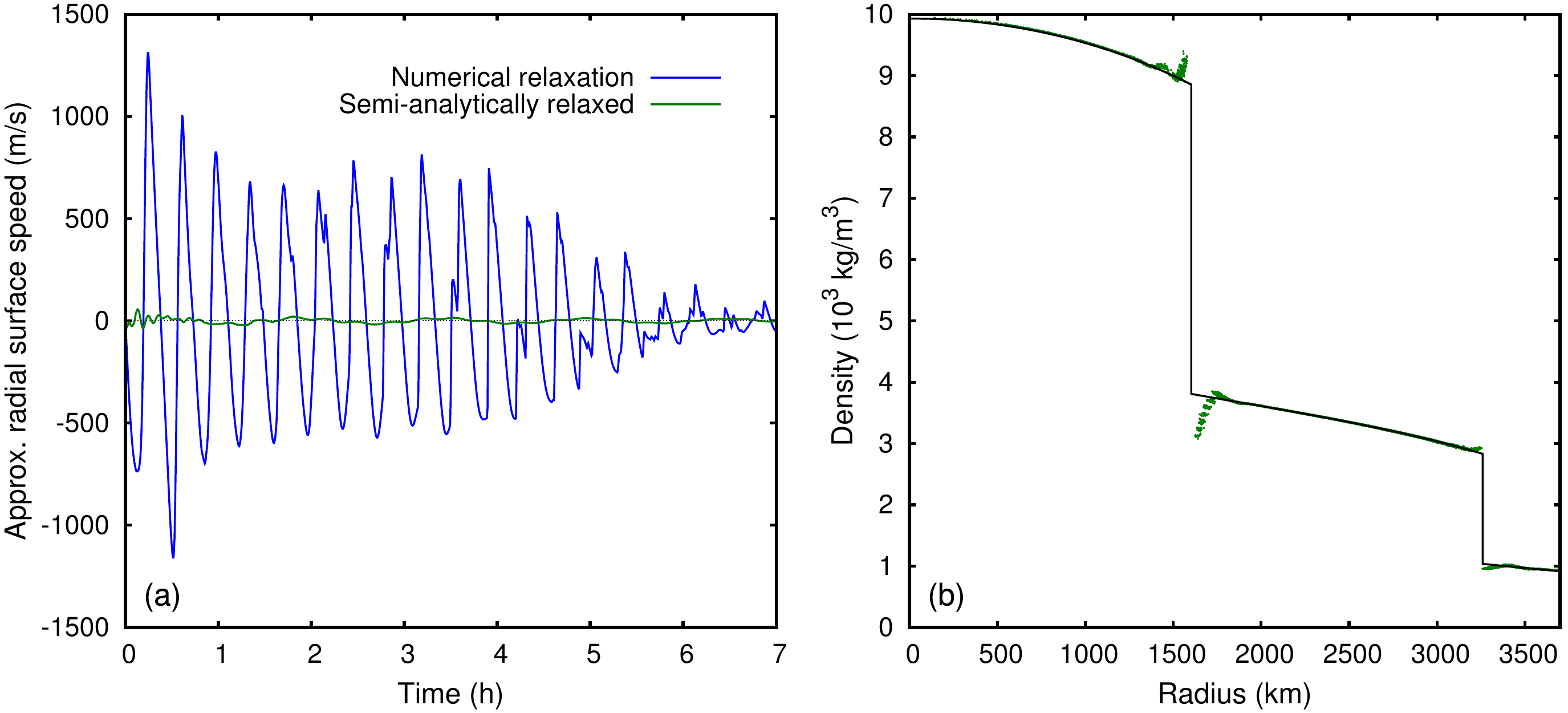}}
	\caption{(a) Comparison of radial surface speeds during numerical relaxation (blue) and for the evolution of an otherwise equal, but initially semi-analytically relaxed body (green) with $\mathrm{M}_\mathrm{Mars}\cong 6.4\times 10^{24}\,\mathrm{kg}$ and the usual 3-layered structure (cf.~Sect.~\ref{sect:scenarios}). (b) Semi-analytically computed density structure (black line) and the result of simulating this body alone, sufficiently long for all remaining fluctuations to settle (green dots). See the online version for color.}
	\label{fig:two_relaxation_plots}
	\end{figure}
Figure~\ref{fig:two_relaxation_plots} (a) illustrates this by comparing radial surface speeds for a numerical relaxation run (i.e.\ starting from homogeneous-density bodies) and a run comprising a body of equal mass and composition -- but set up following our semi-analytical method. While numerical relaxation naturally leads to strong radial oscillations (due to initial gravitational collapse and subsequent rebound), the results of the semi-analytical approach show only some minor residual fluctuations, mainly due to material boundary effects, inherent to SPH \citep[see also][]{Reinhardt2017_Numerical_aspects_of_giant_impact,Woolfson2007_SPH_models_for_major_planets}.
Maximum particle velocities falling below some threshold (relative to the collision velocity) is an often considered criteria for a sufficiently relaxed body \citep[e.g.][and references therein]{Hosono2016_Giant_impact_sims_density_independent_SPH}. 
In the example in Fig.~\ref{fig:two_relaxation_plots} the peak velocity at the onset of numerical relaxation is above 1\,km/s, where the surface escape velocity is about 4.8\,km/s, while the maximum value after semi-analytical relaxation is around 50\,m/s, and drops quickly, mostly still during the initial approach phase of projectile and target. This is only about 1\% of $v_\mathrm{esc}$ (and therefore of a typical collision velocity), where this value is also an often used criterion for a sufficiently relaxed configuration \citep[as e.g.\ in][]{Hosono2016_Giant_impact_sims_density_independent_SPH,Reufer2012_Hit-and-run_Moon_formation}.
The validity of the semi-analytically relaxed initial conditions is also evident from Fig.~\ref{fig:two_relaxation_plots} (b), which illustrates that the calculated profiles are -- except for the typical boundary effects -- very close to equilibrium configurations.

The semi-analytical approach, works almost instantaneously, and practical experience has shown no significant differences to numerically relaxed (but otherwise equal) runs. Thus it is a convenient alternative for producing equilibrated initial conditions in giant collision simulations. The algorithm's implementation in C is freely available to the community upon request\footnote{It was developed for use with our SPH code \emph{miluphCUDA} \citep{Schaefer2016_miluphcuda} -- which is also available to the community -- but can also be adapted to other codes. If you are interested please send an e-mail to \emph{c.burger@univie.ac.at}.}.
In the following we briefly outline the calculation of the hydrostatic structure and the adiabatic internal energy profile.

\subsection*{A.1. Hydrostatic structure calculation}
\label{section_hydrostatic_structure_calculation}
For the given geometry, i.e.\ spherical symmetry, the continuum mechanics problem reduces to simple hydrostatics, even if full solid-body physics is considered in principle.
The Euler representation of the set of equations necessary for describing a spherical, hydrostatic configuration is
	\begin{eqnarray}
	\begin{aligned}
		\frac{\mathrm{d}p(r)}{\mathrm{d}r} &= -\frac{G\,\varrho(r)\,m(r)}{r^2} \\
		\frac{\mathrm{d}m(r)}{\mathrm{d}r} &= 4\pi\,r^2\,\varrho(r) \ ,
		\label{equation_hydrostatic_euler_final_1}
	\end{aligned}
	\end{eqnarray}
with the pressure $p$, density $\varrho$ and integrated (enclosed) mass $m$, all functions of the radial coordinate $r$, and the gravitational constant $G$.
An eos $p = p\,(\varrho, e)$, along with a convenient treatment of the (specific) internal energy $e$, like e.g. $e = e\,(\varrho)$, close the system of equations for $p(r)$, $\varrho(r)$, $m(r)$ and $e(r)$.
The treatment of $e$ is basically open to many different approaches, here its final value is calculated for a given state of adiabatic compression, determined by density alone, neglecting other non-mechanical contributions (see below).

Even though it is possible to directly use these equations, transforming them to their Lagrangian form (with $m$ as the independent variable instead of $r$) is very advantageous when it comes to numerical stability and handiness. This is due to the already initially well-defined domain of $m$-values, and avoidance of negative pressures which can cause serious problems with the eos.
The equivalent of (\ref{equation_hydrostatic_euler_final_1}) then reads
	\begin{eqnarray}
	\begin{aligned}
		\frac{\mathrm{d}r(m)}{\mathrm{d}m} &= \frac{1}{4\pi\,r(m)^2\,\varrho(m)}	\\
		\frac{\mathrm{d}p(m)}{\mathrm{d}m} &= -\frac{G\,m}{4\pi\,r(m)^4}\ .
		\label{equations_hydrostatic_lagrange_final_1}
	\end{aligned}
	\end{eqnarray}
Since these are differential equations for $r$ and $p$, respectively, it is necessary to also calculate $\varrho$ for a given $p$ (and $e$), each time the differential equations' numerical solution is advanced by one step (in $m$).
This can be achieved by inverting the eos, usually given as $p = p\,(\varrho, e)$, to obtain $\varrho = \varrho\,(p, e)$, which is in principle possible as long as the eos is bijective, even though it might not be possible to invert it analytically (as for the Tillotson eos).
Inserting $e\,(\varrho)$ into the eos results in the self-consistency problem $\varrho = \varrho\,\big(p,\,e(\varrho)\big)$.
We solve this problem by a simple fixed-point iteration procedure, following the steps
	\begin{multline}
		p=p(\varrho,\,e)\quad \longrightarrow\quad \varrho = \varrho(p,\,e) = \varrho\big(p,\,e(\varrho)\,\big)\quad \longrightarrow \\ \varrho^{(n+1)}=\varrho\left(p,\,e\left(\frac{\varrho^{(n)}+\varrho^{(n-1)}}{2}\right)\,\right)
		\ .
		\label{equation_hydrostatic_lagrange_final_2}
	\end{multline}
Collecting the pieces together finally allows to calculate $\varrho(m)$, $p(m)$ and $r(m)$ from the differential equations (\ref{equations_hydrostatic_lagrange_final_1}) and from the iteration (\ref{equation_hydrostatic_lagrange_final_2}), and subsequently $e(m) = e\big( \varrho(m) \big)$.

	\begin{figure}
	\centering{\includegraphics[width=250pt]{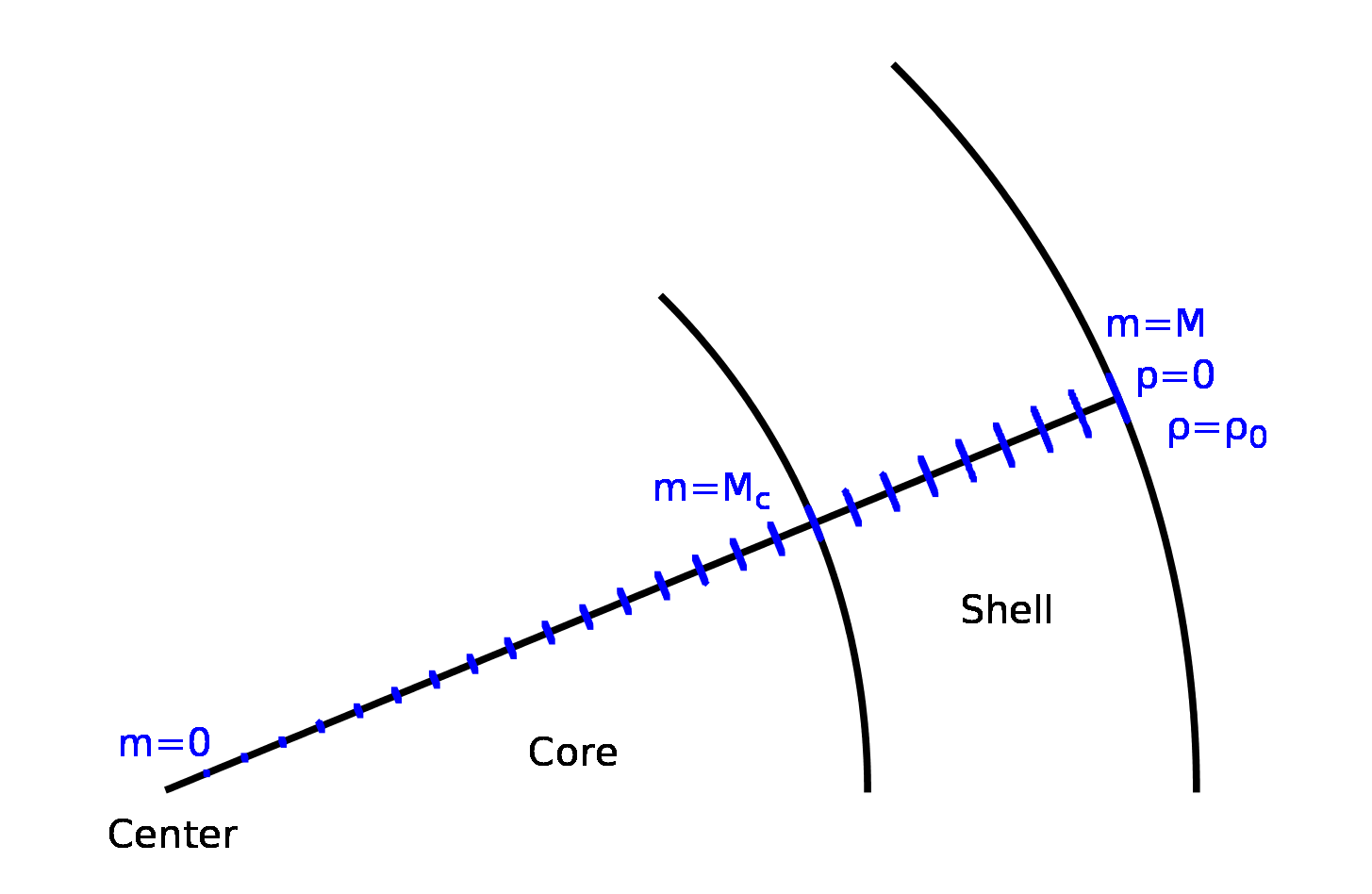}}
	\caption{Schematic illustration of a spherical body with a given core-shell configuration, typical boundary conditions at the outer boundary, and discretization in $m$.}
	\label{fig:mrange_sketch}
	\end{figure}

How this can be carried out practically can be understood with the aid of Fig.~\ref{fig:mrange_sketch}, depicting a body with a core-shell structure.
It is clear that the numerical integration has to start at the outer boundary and be advanced inwards, since initially the properties at the center are usually unknown, while conditions at the outer boundary are specified by the overall mass $M$, a given surface pressure (typically $p=0$) and the density at the surface.
A suitable initial guess for the body's radius $R$ serves as the starting point for the first inwards integration down to the core. In the following an iterative approach varies $R$ until the found solution (between surface and center) converges towards a consistent internal structure, meaning in particular $r(m\!=\!0) \cong 0$ (where the sign of $r(m\!=\!0)$ can be used to test the current $R$ and adjust it for the next iteration).
We found a simple 4th-order Runge Kutta integrator sufficient for this task.

\subsection*{A.2. Internal energy treatment}
\label{section_internal_energy}
The approach we adopted throughout this paper considers no (direct) thermal contributions but only energy stored via adiabatic compression, thus $e=e(\varrho)$.
The time evolution of the specific internal energy due to mechanical work reads
	\begin{equation}
		\frac{\mathrm{d}e}{\mathrm{d}t} = -\frac{p}{\varrho}\, \mathrm{div}(\vec v)\ ,
		\label{equation_treatment_of_e_1}
	\end{equation}
with $\vec v$ being the flow velocity.
We now assume steady, adiabatic compression and integrate this equation to finally obtain $e$ for a given final state (density).
The basic assumption in the following steps is homogeneous compression, i.e.\ homogeneous density and internal energy in a sufficiently big volume surrounding the point under consideration, for every arbitrary but fixed instant of time.
Using this, the substantial derivative in (\ref{equation_treatment_of_e_1}), $\mathrm{d}e/\mathrm{d}t = \partial e/\partial t + \vec v \, \mathrm{grad}(e)$, reduces to $\partial e/\partial t$, since $\mathrm{grad}(e)$ vanishes.
Assuming a constant volume-flow towards all points in the considered volume,\ $-\mathrm{div}(\vec v) = \mathrm{const.} = \alpha$, \ leads to an adapted version of (\ref{equation_treatment_of_e_1}):
	\begin{equation}
		\frac{\partial e}{\partial t} = \alpha\, \frac{p}{\varrho} \ .
		\label{equation_treatment_of_e_3}
	\end{equation}
The next step is to invoke the continuity equation, $\partial\varrho/\partial t + \mathrm{div}(\varrho\vec v) = 0$, where the divergence can be split (\,$\mathrm{div}(\varrho\vec v) = \vec v\, \mathrm{grad}(\varrho) + \varrho\, \mathrm{div}(\vec v)$\,) and the density gradient vanishes, leading to $\partial\varrho/\partial t = \alpha \varrho$, a simple ordinary differential equation with the solution
	\begin{equation}
		\varrho (t) = \varrho_0\, \mathrm{e}^{\alpha t} \ ,
		\label{equation_treatment_of_e_6}
	\end{equation}
where $\varrho_0$ represents the uncompressed density.
Equation (\ref{equation_treatment_of_e_6}) simply describes the development of the density for homogeneous compression, where $\alpha$ determines the (constant) volume-flow into the considered region.
Inserting (\ref{equation_treatment_of_e_6}) into (\ref{equation_treatment_of_e_3}), and setting (arbitrarily) $\alpha = 1$ \,for simplicity, eventually leads to
	\begin{equation}
		\frac{\partial e}{\partial t} = \frac{p}{\varrho_0}\,\mathrm{e}^{-t} \ .
		\label{equation_treatment_of_e_7}
	\end{equation}
Together with the chosen eos and the analytical result for $\varrho(t)$ in (\ref{equation_treatment_of_e_6}), the pressure can finally be expressed as $p = p(\varrho,\,e) = p\big( \varrho(t),\, e \big)$.
Inserting this into (\ref{equation_treatment_of_e_7}) it becomes an ordinary differential equation for the time evolution of $e(t)$, albeit the right-hand side might be complicated, depending on the choice of eos.
To finally end up with $e(\varrho)$, this equation can be numerically integrated, starting at $t=0$, with $e(0) = 0$, where the upper integration limit $t_{\mathrm{end}}$ is found by inverting (\ref{equation_treatment_of_e_6}) for the desired $\varrho$, i.e.\ for exactly the respective state of compression, $t_{\mathrm{end}} = \ln( \varrho/\varrho_0 )$.
}

\section*{Appendix B -- Collision outcome data}
Summarized data on the outcome of all hit-and-run collision scenarios is provided in Tab.~\ref{tab:hit-and-run_results}. Data on all computed head-on scenarios can be found in Tab.~\ref{tab:head-on}.

\begin{table}
\centering
\scriptsize
\begin{threeparttable}
	\caption{Summarized data on all hit-and-run scenarios.}
	\label{tab:hit-and-run_results}
	\begin{tabular}[b]{p{1.0cm} p{0.7cm} p{0.5cm} p{0.6cm} C{0.4cm} C{0.4cm} C{0.7cm} C{0.75cm} C{1.4cm} C{1.4cm}}
	\hline
	\hline
	log($M_\mathrm{tot}$)	&	$v/v_\mathrm{esc}$	&	$\alpha (^\circ)$	&	$\gamma$	&	$f_\mathrm{lf}$	&	$f_\mathrm{slf}$	&	$wmf_\mathrm{lf}$	&	$wmf_\mathrm{slf}$	&	$wmf_\mathrm{lf\leftarrow proj}$	&	$wmf_\mathrm{slf\leftarrow targ}$	\\
	\hline
\ \ \ \ \ \,22 & 2.5 & 45 & 0.05 &   95 &    1.9 &    9.4 &    7.9 &    0.13 &     1.6 \\
\ \ \ \ \ \,22 & 2.5 & 45 & 0.11 &   90 &    6.4 &    9.4 &    7.6 &    0.22 &     1.5 \\
\ \ \ \ \ \,22 & 2.5 & 45 & 0.5 &   65 &     30 &      9 &    8.6 &    0.63 &     1.1 \\
\hline
\ \ \ \ \ \,23 & 1.5 & 30 & 0.05 &   99 &  0.027 &    9.5 &    9.8 &    0.36 &     2.5 \\
\ \ \ \ \ \,23 & 1.5 & 30 & 0.11 &   97 &   0.11 &    9.2 &     15 &    0.69 &     0.9 \\
\ \ \ \ \ \,23 & 1.5 & 30 & 0.2 &   95 &    1.2 &      9 &     14 &     1.2 &     2.6 \\
\ \ \ \ \ \,23 & 1.5 & 30 & 0.5 &   68 &     31 &    8.8 &    8.9 &     1.5 &     3.1 \\
\ \ \ \ \ \,23 & 1.5 & 45 & 0.11 &   91 &    7.5 &    9.5 &    9.6 &    0.38 &     3.5 \\
\ \ \ \ \ \,23 & 1.5 & 45 & 0.5 &   68 &     32 &    9.6 &    9.5 &     1.1 &     2.2 \\
\ \ \ \ \ \,23 & 1.5 & 60 & 0.11 &   90 &    9.5 &    9.9 &    9.3 &    0.18 &     1.5 \\
\ \ \ \ \ \,23 & 1.5 & 60 & 0.5 &   67 &     33 &    9.9 &    9.8 &    0.48 &    0.89 \\
\ \ \ \ \ \,23 & 1.5 & 75 & 0.11 &   90 &    9.9 &     10 &    9.4 &   0.057 &     0.2 \\
\ \ \ \ \ \,23 & 1.5 & 90 & 0.11 &   90 &     10 &     10 &    9.6 &   0.018 &   0.013 \\
\ \ \ \ \ \,23 & 2.5 & 30 & 0.05 &   96 &   0.16 &    8.9 &    8.4 &    0.21 &     4.4 \\
\ \ \ \ \ \,23 & 2.5 & 30 & 0.11 &   89 &   0.62 &    8.5 &    7.2 &    0.32 &     2.8 \\
\ \ \ \ \ \,23 & 2.5 & 30 & 0.2 &   81 &    6.6 &    8.3 &    6.1 &    0.48 &       2 \\
\ \ \ \ \ \,23 & 2.5 & 30 & 0.5 &   62 &     22 &    7.8 &    6.4 &    0.89 &     1.7 \\
\ \ \ \ \ \,23 & 2.5 & 45 & 0.02 &   98 &   0.22 &    9.6 &    6.4 &   0.053 &    0.61 \\
\ \ \ \ \ \,23 & 2.5 & 45 & 0.05 &   95 &    1.9 &    9.4 &    5.6 &    0.12 &    0.61 \\
\ \ \ \ \ \,23 & 2.5 & 45 & 0.11 &   90 &    6.2 &    9.3 &    6  &    0.21 &    0.73 \\
\ \ \ \ $^\mathrm{d}$23 & 2.5 & 45 & 0.11 & 90 & 4.9 & 0.24 & 5.1 & 0.24 & -- \\
\ \ \ \ \,$^\mathrm{s}$23 & 2.5 & 45 & 0.11 & 90 & 7  & 9.3 & 6.8 & 0.15 & 0.45 \\
$^\mathrm{0.3M}$23 & 2.5 & 45 & 0.11 & 90 & 6.1 & 9.3 & 6  & 0.23 & 0.85 \\
\ \ $^\mathrm{1M}$23 & 2.5 & 45 & 0.11 & 90 & 6.1 & 9.3 & 5.7 & 0.23 & 0.91 \\
\ \ \ \ \ \;23 & 2.5 & 45 & 0.2 &   83 &     13 &    9.2 &    7.2 &    0.33 &       1 \\
\ \ \ \ \;\!$^\mathrm{a}$23 & 2.5 & 45 & 0.5 & 65 & 29 & 4.4 & 4  & 0.34 & 0.52  \\
\ \ \ \ \ \;23 & 2.5 & 45 & 0.5 &   65 &     30 &    8.9 &      8 &     0.6 &    0.89 \\
\ \ \ \ \;\!$^\mathrm{b}$23 & 2.5 & 45 & 0.5 & 65 & 31 & 18.5 & 16 & 0.92 & 1.4 \\
\ \ \ \ \:\,\!$^\mathrm{s}$23 & 2.5 & 45 & 0.5  & 66 & 31 & 9.1 & 8.4 & 0.38 & 0.62 \\
$^\mathrm{0.3M}$23 & 2.5 & 45 & 0.5  & 65 & 30  & 8.9 & 8.1 & 0.64 & 0.96 \\
\ \ $^\mathrm{1M}$23 & 2.5 & 45 & 0.5 & 65 & 30 & 8.9 & 8.1 & 0.64 & 1 \\
\ \ \ \ \ \,23 & 2.5 & 60 & 0.02 &   98 &    1.2 &    9.9 &    5.6 &   0.024 &    0.12 \\
\ \ \ \ \ \,23 & 2.5 & 60 & 0.05 &   95 &      4 &    9.8 &    7.3 &   0.048 &    0.44 \\
\ \ \ \ \ \,23 & 2.5 & 60 & 0.11 &   90 &    9.4 &    9.8 &    8.1 &   0.084 &    0.49 \\
\ \ \ \ $^\mathrm{d}$23 & 2.5 & 60 & 0.11 & 90 & 9   & 0.11 & 7.6 & 0.11 & -- \\
\ \ \ \ \ \,23 & 2.5 & 60 & 0.5 &   66 &     33 &    9.7 &    9.2 &    0.21 &    0.35 \\
\ \ \ \ \ \,23 & 3.5 & 30 & 0.11 &   85 &    0.1 &      8 &    7.1 &    0.22 &     2.7 \\
\ \ \ \ \ \,23 & 3.5 & 30 & 0.5 &   55 &     18 &    6.8 &    5.2 &    0.56 &     1.1 \\
\ \ \ \ \ \,23 & 3.5 & 45 & 0.02 &   98 &   0.02 &    9.4 &    1.6 &    0.03 &       0 \\
\ \ \ \ \ \,23 & 3.5 & 45 & 0.05 &   95 &    1.4 &    9.2 &      4 &    0.07 &    0.12 \\
\ \ \ \ \ \,23 & 3.5 & 45 & 0.11 &   89 &    5.1 &    9.1 &    4.5 &    0.14 &    0.23 \\
\ \ \ \ \ \,23 & 3.5 & 45 & 0.5 &   64 &     29 &    8.5 &    7.2 &    0.36 &    0.39 \\
\ \ \ \ \ \,23 & 3.5 & 60 & 0.02 &   98 &   0.95 &    9.8 &    5.1 &   0.014 &   0.035 \\
\ \ \ \ \ \,23 & 3.5 & 60 & 0.11 &   90 &    9.2 &    9.7 &    7.7 &   0.048 &     0.2 \\
\ \ \ \ \ \,23 & 3.5 & 60 & 0.5 &   66 &     33 &    9.5 &    8.9 &   0.092 &    0.16 \\
\ \ \ \ \ \,23 & 4.5 & 30 & 0.11 &   81 &  0.053 &    7.2 &    5.7 &    0.14 &     3.2 \\
\ \ \ \ \ \,23 & 4.5 & 30 & 0.5 &   46 &     12 &    5.1 &    2.7 &    0.35 &    0.79 \\
\ \ \ \ \ \,23 & 4.5 & 45 & 0.11 &   88 &    4.2 &    8.8 &    2.9 &   0.075 &    0.09 \\
\ \ \ \ \ \,23 & 4.5 & 45 & 0.5 &   63 &     27 &      8 &    6.4 &    0.16 &    0.21 \\
\ \ \ \ \ \,23 & 4.5 & 60 & 0.11 &   90 &    8.9 &    9.6 &    7.3 &   0.027 &    0.08 \\
\ \ \ \ \ \,23 & 4.5 & 60 & 0.5 &   66 &     33 &    9.4 &    8.7 &   0.052 &   0.085 \\
\ \ \ \ \ \,23 & 5.5 & 45 & 0.11 &   87 &      3 &    8.5 &    0.2 &   0.042 &     0.1 \\
\ \ \ \ \ \,23 & 5.5 & 45 & 0.5 &   61 &     24 &    7.4 &    5.1 &   0.084 &    0.11 \\
\hline
\ \ \ \ \ \,24 & 1.5 & 90 & 0.11 &   90 &     10 &     10 &    9.8 &   0.009 &       0 \\
\ \ \ \ \ \,24 & 2.5 & 30 & 0.11 &   89 &   0.49 &    8.5 &    6.5 &    0.36 &     3.7 \\
\ \ \ \ \ \,24 & 2.5 & 45 & 0.05 &   95 &    1.9 &    9.6 &    3.5 &    0.12 &    0.27 \\
\ \ \ \ \ \,24 & 2.5 & 45 & 0.11 &   90 &    6.2 &    9.4 &    5.2 &    0.21 &    0.46 \\
\ \ \ \ \ \,24 & 2.5 & 45 & 0.5 &   66 &     31 &    9.1 &      8 &    0.55 &    0.69 \\
\hline
\ \ \ \ \ \,25 & 1.5 & 45 & 0.11 &   91 &    7.7 &    9.6 &    7.4 &    0.38 &     1.8 \\
\ \ \ \ \ \,25 & 1.5 & 90 & 0.11 &   90 &     10 &    9.9 &    9.9 &  0.0015 &       0 \\
\ \ \ \ \ \,25 & 2.5 & 30 & 0.11 &   89 &    0.2 &    8.3 &    1.9 &    0.37 &     1.5 \\
\ \ \ \ \ \,25 & 2.5 & 45 & 0.05 &   95 &    1.4 &    9.5 &  0.096 &   0.088 &   0.096 \\
\ \ \ \ \ \,25 & 2.5 & 45 & 0.11 &   90 &    6.7 &    9.3 &    2.7 &    0.15 &    0.15 \\
$^\mathrm{0.3M}$25 & 2.5 & 45 & 0.11 & 90 & 6.8 & 9.4 & 2.3 & 0.17 & 0.14 \\
\ \ $^\mathrm{1M}$25 & 2.5 & 45 & 0.11 & 90 & 6.9 & 9.5 & 2.1 & 0.18 & 0.13 \\
$\!\!\! ^\mathrm{2.25M}$25 & 2.5 & 45 & 0.11 & 90 & 7.2 & 9.5 & 2 & 0.18 & 0.12 \\
\ \ \ \ \ \,25 & 2.5 & 45 & 0.5 &   66 &     31 &      9 &    7.9 &    0.39 &    0.48 \\
\ \ \ \ \ \,25 & 3.5 & 45 & 0.11 &   89 &    4.8 &    9.1 &  0.039 &   0.065 &   0.039 \\
	\hline
	\end{tabular}
	\begin{tablenotes}
	\item {\bf Notes.} Fractions ($f$) for the largest fragment (lf) and second-largest fragment (slf) are relative to $M_\mathrm{tot}$. Water mass fractions ($wmf$) are relative to fragment masses, and arrows ($\leftarrow$) denote \emph{originating from} to indicate transfer of water. All fractions are given in \%. $^\mathrm{d}$are dry-target runs (see~Sect.~\ref{sect:water_amount_distribution}). $^\mathrm{a, b}$are runs with initial wmf of 5\% and 20\%, respectively (Sect.~\ref{sect:water_amount_distribution}). $^\mathrm{s}$include material strength (Sect.~\ref{sect:material_strength}). $^\mathrm{0.3M, 1M, 2.25M}$indicate high resolution with 0.3, 1 and 2.25 million particles (Sect.~\ref{sect:resolution}).
	\end{tablenotes}
	\end{threeparttable}
	\end{table}


	\begin{table}
	\centering
	\scriptsize
	\begin{threeparttable}
	\caption{Summarized data on all head-on scenarios.}
	\label{tab:head-on}
	\begin{tabular}[b]{p{1.0cm} p{0.7cm} p{0.5cm} p{0.6cm} C{0.4cm} C{0.7cm} C{0.7cm} C{1.4cm}}
	\hline
	\hline
	log($M_\mathrm{tot}$)	&	$v/v_\mathrm{esc}$	&	$\alpha (^\circ)$	&	$\gamma$	&	$f_\mathrm{lf}$	&	$f_\mathrm{slf}$	&	$wmf_\mathrm{lf}$	&	$wmf_\mathrm{lf\leftarrow proj}$	\\
	\hline
23 & 1.5 & 0 & 0.11 &   99 & 0.0016 &    9.2 &    0.77 \\
23 & 1.5 & 0 & 0.25 &   97 &   0.05 &    8.3 &     1.5 \\
23 & 1.5 & 0 & 0.5 &   96 &  0.036 &      8 &     2.5 \\
23 & 2.5 & 0 & 0.11 &   94 &  0.041 &    8.1 &    0.59 \\
23 & 2.5 & 0 & 0.25 &   83 &  0.099 &    6.8 &     1.1 \\
23 & 2.5 & 0 & 0.5 &   69 &    0.3 &    5.7 &     1.6 \\
23 & 3.5 & 0 & 0.11 &   82 &   0.14 &    7.1 &    0.47 \\
23 & 3.5 & 0 & 0.25 &   52 &   0.62 &    5.2 &    0.71 \\
23 & 3.5 & 0 & 0.5 &  4.6 &      5 &    4.3 &     1.2 \\
23 & 4.5 & 0 & 0.11 &   53 &    1.4 &    1.9 &    0.32 \\
23 & 4.5 & 0 & 0.5 &  1.4 &    1.2 &    2.7 &     1.2 \\
\hline
24 & 1.5 & 0 & 0.25 &   98 &  0.017 &    8.5 &     1.5 \\
24 & 2.5 & 0 & 0.25 &   81 &   0.27 &    6.6 &     1.2 \\	
	\hline
	\end{tabular}
	\begin{tablenotes}
	\item {\bf Notes.} Fractions ($f$) for the largest fragment (lf) and second-largest fragment (slf) are relative to $M_\mathrm{tot}$. Water mass fractions ($wmf$) are relative to fragment masses, and the arrow ($\leftarrow$) denotes \emph{originating from} to indicate transfer of water. All fractions are given in \%.
	\end{tablenotes}
	\end{threeparttable}
	\end{table}

\bibliographystyle{spbasic.bst}
\bibliography{references_scaling.bib,references_small_bodies.bib,references_atmospheres.bib,references_strength.bib,references_hit-and-run.bib,references_mine.bib,references_EOS.bib,references_SPH.bib,references_Planet_formation.bib,references_Water.bib}


%

%



\end{document}